\documentclass[aps,prb,twocolumn,amsmath,longbibliography,amssymb,superscriptaddress,10pt]{revtex4-2}
\usepackage[margin=1in]{geometry}
\usepackage[english]{babel}
\usepackage{etoolbox}
\usepackage[shortlabels]{enumitem}
\usepackage[scr=boondox]{mathalpha}   
\usepackage{booktabs}

\usepackage{amsmath,amsthm,amssymb,amsfonts}
\usepackage{stmaryrd}
\usepackage{accents}
\usepackage{bm}
\usepackage{leftindex}
\usepackage{dsfont} 
\usepackage[linkcolor=blue,colorlinks=true]{hyperref}
\usepackage{graphicx}
\usepackage{subcaption,caption}
\usepackage{tabularx,ragged2e}
\newcolumntype{C}{>{\centering\arraybackslash}X}

\graphicspath{
    {figures/}
}

\docsvlist{A,B,C,D,E,F,G,H,I,J,K,L,M,N,O,P,Q,R,S,T,U,V,W,X,Y,Z}

\docsvlist{A,B,C,D,E,F,G,H,I,J,K,L,M,N,O,P,Q,R,S,T,U,V,W,X,Y,Z}

\docsvlist{j}

\docsvlist{R,k,p,q,r,g,x,y}

\def \ve {\varepsilon}

\def \ket {\rangle}

\def \sro {Sr$_2$RuO$_4$}

\newcommand{\bulk}[1][1]{j_{#1\text{b}}}
\newcommand{\edge}[1][1]{j_{#1\text{e}}}
\newcommand{\bulkunit}[1][1]{\sJ_{#1\text{b}}}
\newcommand{\edgeunit}[1][1]{\sJ_{#1\text{e}}}

\usepackage{xcolor}
\usepackage{transparent}

\newif\ifshow
\showtrue

\newcommand\revise[1]{\ifshow{#1}\else{\color{brown}#1}\fi}

\begin{document}


\title{Phase sensitive information from a planar Josephson junction}
\author{Andrew C. Yuan}
\email[]{andrewcyuan@gmail.com}
\affiliation{Department of Physics, Stanford University, Stanford, CA 94305, USA}

\author{Steven A. Kivelson}
\affiliation{Department of Physics, Stanford University, Stanford, CA 94305, USA}
\begin{abstract}
\revise{
Josephson tunneling across a planar junction generally depends on the relative twist angle, $\theta$, between the  two layers. 
However, if under a discrete rotation, the order parameter in one layer is odd and the other is even (as, e.g., for a  $s$-wave to $d_{x^2-y^2}$-wave junction under a $\pi/2$ rotation)  then the bulk Josephson current 
vanishes for all $\theta$. 
Even in this case, we show that for a finite junction, the Josephson current, $J$, has a nonzero edge contribution that depends on $\theta$ and the orientation of the junction edges in ways that can serve as an unambiguous probe of the order parameter symmetry
of any time-reversal preserving system (including multiband  systems and those in which  spin-orbit coupling is significant). 
We also analyze the microscopic considerations that determine the magnitude of $J$.
}
\end{abstract}
\date{\today}

\maketitle
\section{Introduction}

Famously, phase sensitive measurements \cite{neils2002experimental,tsuei2000pairing,PhysRevLett.73.593,PhysRevLett.74.4523,tanaka1996theory,tanaka1997theory,kashiwaya2000tunnelling} 
of the superconducting order in the cuprate high temperature superconductors resolved ongoing debates concerning the symmetry of the order parameter, establishing the key fact that these are $d$-wave superconductors.  
Most of the material systems currently being studied for which similar issues arise are either highly layered (i.e. quasi-two dimensional), like the cuprates, or else are explicitly two dimensional (2D), such as various explicitly two-dimensional layered structures made from van-der Waals materials, especially graphene.  
Edge junctions of the sort used in the original cuprate experiments are thus often difficult to engineer and sometimes difficult to interpret given the complex nature of such edges.  
Conversely, many quasi-2D materials cleave relatively easily such that the normal to the surface (hence-forth the ``$z$'' direction) is the least conducting direction. 
This geometric consideration is still clearer in the case of 2D materials.

With this in mind, we analyze several methods (each easily generalizable) by which $z$-direction planar Josephson junctions can be constructed to extract phase sensitive information concerning the symmetry of the  order parameter in a candidate unconventional superconductor (SC). (This is somewhat in the spirit of early experiments \cite{li1999bi} 
on the cuprates using twist-junctions, which only now are showing the expected behaviors \cite{zhao2023time,zhu2021presence}.) 
Indeed, with somewhat more engineering, the same sort of analysis can be extended to extract information concerning the magnitude of the superconducting gap as a function of position along the Fermi surface (FS).

\begin{figure}[ht]
\centering
\includegraphics[width=.96\columnwidth]{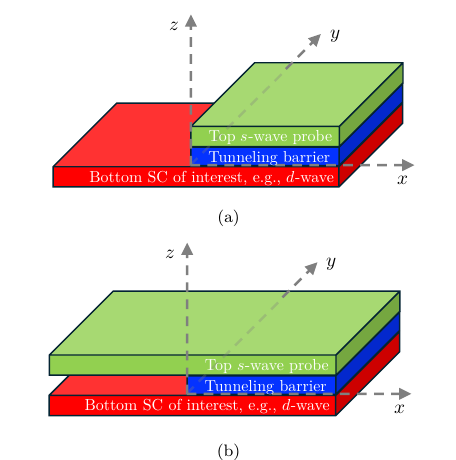}
\caption{Junction geometry.  (a) shows the physical geometry where the upper SC terminates at an edge at $x=0$.  (b) shows the model geometry used to simplify microscopic calculations, in which both SCs are infinite, but tunnelling between them is only permitted for $x>0$.
}
\label{fig:setup}
\end{figure}

\begin{figure}[ht]
\centering
\includegraphics[width=0.6\columnwidth]{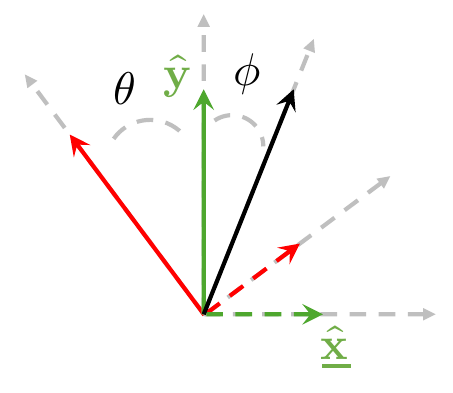}
\caption{\revise{Angles. The green arrows denote the top SC; $\hat{\by}$ is a given \textit{principle axis}, while $\hat{\underline{\bx}}$ is the corresponding reciprocal vector such that $\hat{\underline{\bx}}\perp \hat{\by}$ (note that the two primitive vectors $\hat{\bx},\hat{\by}$ are not necessarily perpendicular, e.g., hexagonal lattice).
The black arrow denotes the edge so that $\phi$ is the angle of the edge relative to the principle axis of the upper SC.
The red arrows denote the bottom SC so that $\theta$ is relative twist between the principles axes of the top and bottom SC and that the edge is at angle $\theta+\phi$ relative to the principle axis of the lower SC.}}
\label{fig:angles}
\end{figure}

We consider a planar Josephson junction with the geometry shown in Fig. \ref{fig:setup}a. 
The top portion of the junction is a half-plane of what we will assume is a probe SC whose gap symmetry is well-established, and the bottom portion is a full plane of an interesting SC, whose order parameter symmetry and gap structure we would like to determine.  
There are two angles that define the geometry of the junction: $\theta$, the relative twist angle between the \textit{principle axes} 
of the top and bottom SCs, and $\phi$, the angle of the boundary of the top SC relative to the principle axis of the top SC (see Fig. \ref{fig:angles}).
We consider a situation in which  the magnitude of the single particle tunnelling between the two superconductors, $t_\perp$, is weak - i.e. this is assumed to be a planar version of an SIS junction.  
This means that the energy $J(\alpha)$ can be expressed as a function of $\alpha$, the difference in the SC phase across the junction.
\revise{If time-reversal symmetry (TRS) is unbroken, the energy is invariant under $\alpha \mapsto -\alpha$} and thus can be expressed as a series of cosines,
\begin{align}
    J(\alpha) = -J_0-J_1\cos(\alpha) -J_2\cos(2\alpha) + \cdots
\end{align}
where $J_1 \sim t_\perp^2$ plus corrections of order $t_\perp^6$, while $J_2 \sim t_\perp^4 + \cdots$.  
Moreover, each $J_n$ has a bulk contribution, proportional to the area $A$ of the junction, an edge contribution proportional to the length of the edge, $L$, and smaller terms that depend on further details of the shape of the junction.  Focusing on the lowest harmonic, we can express $J_1$ as
\begin{align}
    J_1 = \bulk(\theta, \bg) \frac{A}{a^2} + \edge(\theta,\phi, \bg) \frac{L}{a} + \cdots
\end{align}
where $a$ is the lattice constant (which will be set $=1$ in the main text), $\bulk/a^2$ is the bulk Josephson coupling per unit area, $\edge/a$  is the edge contribution  per unit length of edge, and $\cdots$ signifies terms that are independent of the size of the junction.
Here we have included explicit factors of $a^n$ so that both $\bulk$ and $\edge$ have the same units and depend on a set of properties $\bg$ (which will often be left implicit) that characterize the bulk superconducting state of the two superconductors; the bulk $\bulk$ (and higher orders $\bulk[2],...$) only depends on the twist angle $\theta$ since it is oblivious of the edge, while the edge contribution $\edge$ depends on both $\theta$ and $\phi$. 

\revise{
The article is organized as follows. In the ``Results'' section, we first summarize the general symmetry considerations of the bulk and edge contributions, which can be used to probe the SC order parameter symmetry.
In particular, we will be primarily focused on the explicit example of a single band singlet pairing (SBSP) ($D_4$ symmetric) $s$-$d_{x^2-y^2}$ wave junction, though we emphasize that the symmetry considerations are completely general and apply to any TRS preserving system (including multiband systems and those in which spin-orbit coupling is significant).
These symmetry properties are summarized in Table \eqref{tab:edge}.
Next, we summarize the microscopic considerations that determine the scaling of the bulk and edge contributions.
By restricting our attention to SBSP junctions, our argument reveals that the scaling is determined mostly by momenta near the normal state FS and thus we expect similar arguments to hold for general multi-band systems \footnote{There may, however, be special circumstances where at the relevant points on the Fermi surface, the structure of the Bloch states in the two systems could be sufficiently distinct - reflecting some non-trivial aspect of their multi-orbital origin - that this could affect the results.}.
Next, we summarize that the presence of disorder does not affect the previously discussed symmetry considerations. 
Even the scaling of the quantities $\bulk,\edge$ is typically not greatly affected so long as the correlation length is substantial.
Finally, we provide explicit details including analytic formulas for the restricted case of SBSP junctions and numerical calculations of the SBSP $s$-$d_{x^2-y^2}$ wave junction to support our conclusions regarding symmetry and scaling.
In the ``Discussion'' section, we propose an experimental recipe and discuss potential scenarios in which our setup may provide useful information.
}
\section{Results}

\subsection{Symmetry Considerations}

\begin{table}[ht!]
\begin{subtable}{1\columnwidth}
\centering
\begin{tabularx}{1\columnwidth}{C C C}
\toprule
$\theta \mapsto C_{2n}\theta$ & $\phi\mapsto C_{2n}\phi$ & $\theta,\phi\mapsto\sM\theta,\sM\phi$\\
\midrule
$r^-$ & $r$ & $m$\\
\bottomrule
\end{tabularx}
\caption{Symmetries of $\edge$}
\end{subtable}
\\[4ex]
\begin{subtable}{1\columnwidth}
\begin{tabularx}{1\columnwidth}{C C C C}
\toprule
$(0,0)$ &$\displaystyle\left(\frac{\pi}{2n},0\right)$ & $\displaystyle\left(0,\frac{\pi}{2n}\right)$ & $\displaystyle\left(\frac{\pi}{2n},\frac{\pi}{2n}\right)$ \\ [1.5ex]
\midrule
$m$ & $m r^+$ & $m r^-$ & $m r$ \\
\bottomrule
\end{tabularx}
\caption{Zeros of $\edge$}
\end{subtable}
\caption{$D_{2n}$ point group. Let $C_{2n},\sM$ denote $2n$-fold rotation and mirror symmetry along the edge $\hat{\by} \mapsto -\hat{\by}$ so that the SC gaps $\Delta^\pm \mapsto r^\pm \Delta^\pm, m^\pm \Delta^\pm$ (where $r^\pm,m^\pm \in \{\pm 1\}$) under $C_{2n},\sM$, respectively.  Write $r=r^+r^-$ and $m=m^+m^-$. 
(For example, in a tetragonal $s$-$g$ junction, $r^\pm = 1$ since $s,g$-waves are even under $C_4$ rotation, while $m^\pm=\pm 1$ since an $s,g$-wave is even/odd under mirror symmetry $\sM$, respectively.)
(a). The bottom row shows the symmetry of $\edge$ under transform given by the top row, e.g., $\edge \mapsto r^- \edge$ under $\theta \mapsto C_{2n}\theta$ while $\phi$ is kept invariant. In particular, $\edge$ depends on $\theta,\phi$ only modulo $\pi/n$. (b). The top row $(\phi,\theta +\phi) \mod \pi/n$ shows the angles that are highly symmetric due to the symmetries in (a). In particular, the top row is a zero of $\edge$ if the bottom row $=-1$. See Supplementary Note 1 for proof.
}
\label{tab:edge}
\end{table}

If at least one SC has non $s$-wave symmetry, then the $\theta$ dependence of the bulk $\bulk$ is strong such that it vanishes under some circumstances. 
For instance, if the $s$-wave SC is orthorhombic and the $d$-wave SC is tetragonal, then $\bulk$ is generically non-zero \cite{kleiner1996pair}, but vanishes for any $\theta$ for which the orthrhombic axis aligns with the gap-nodes of the $d$-wave SC. 
(Similar symmetry considerations apply to all odd bulk harmonics, $\bulk[2n+1,]$, while the expected angle dependence of even bulk harmonics $\bulk[2n,]$ is much weaker and less revealing.)  
Therefore, the symmetry considerations of $\bulk$ can generally be used to probe the SC order parameter symmetry (e.g., $d$-$d$ wave junction \cite{zhao2023time}).
However, if under a discrete rotation, the order parameter in one layer is odd and the other is even, then the bulk Josephson current of odd harmonics vanishes for all $\theta$ (see Eq. \eqref{eq:bulk-vanishes}).

As an illustrative example, we shall consider the case where the top and bottom layers are (SBSP $D_4$ irreps) $s$- and $d_{x^2-y^2}$-wave SCs, respectively.
In this case, although $\bulk(\theta)=0$ for all twist angles $\theta$, the $C_4$ symmetry is broken by the edge, which implies a non-zero $\edge$ with a nontrivial $\theta$ and $\phi$ dependence. 
In particular (see Sec. \eqref{sec:general-rotation}),
\begin{enumerate}[label=\textbf{\arabic*)}]
    \item $\edge \mapsto -\edge$ under $\theta\mapsto \theta +\pi/2$ 
    \item $\edge \mapsto -\edge$ under $\phi \mapsto \phi +\pi/2$ 
    \item $\edge \mapsto \edge$ under $\theta, \phi \mapsto -\theta,-\phi$
    \label{claim:symmetry}
\end{enumerate}
\revise{
Importantly, the symmetry of an $s$-$d$ junction implies that 
(1) $\edge$ depends on $(\theta,\phi)$ only modulo $\pi/2$ (albeit with a sign change). 
In particular, for any fixed $\theta$, 
$\edge$ vanishes at some critical edge-angle $\phi_c(\theta)$ across which $\edge$ changes sign  \footnote{Equally, this can be expressed as a critical twist angle, $\theta_c(\phi)$, for given edge angle $\phi$.}.  (2) Certain angles $(\phi,\theta)$ are highly symmetric.
For example, when the edge is oriented along a symmetry axis, i.e. when $\phi =0$ or $\pi/4 \mod \pi/2$, $\edge$ vanishes whenever $\theta + \phi =\pi/4 \mod \pi/2$. 
Other order parameter symmetries would lead to different patterns of critical angles as tabulated in Table. \eqref{tab:edge}.
}

\subsection{Microscopic (quantitative) considerations}

The explicit calculations for model electronic band-structures, discussed below, reveal a number of  more detailed, microscopic features of the two SCs that implicitly affect the Josephson coupling.  
Particularly important are the magnitude of the two gaps, the structure of the Fermi surfaces and the dependence of the gap function on position along the Fermi surface.  
Understanding these is essential for making quantitative estimates of the various contributions to $J_1$, important for the practical design of such experiments.  Conversely, these features can also be exploited to infer more microscopic properties of the SC state.

Of these, the most critical are those that govern the maximum magnitudes of $\bulk$ and $\edge$. 
In cases in which bulk coupling is symmetry allowed, the dependence of $\bulk$ on the SC gap magnitude $|\Delta|$ is expressible as (see Eq. \eqref{eq:bulk-contr} and Fig. \ref{fig:FS}a,b)
\begin{align}
     \bulk =\bulkunit\times \frac{t_\perp^2 |\Delta|}{E_F^2}\times\left(\frac {|\Delta|} {E_F}\right)^{\delta}
     \label{eq:bulk-scale}
\end{align}
where $\bulkunit$ is a dimensionless quantity  of order one,   $E_F$ is an appropriate average of the Fermi energies in the two SCs, and the exponent $\delta=0$ in the case in which the FSs of the two SCs intersect at discrete points in the Brillouin Zone ($\dB\dZ$), and $\delta=1$ when they do not intersect.
(Two FSs can be considered \textit{non-intersecting}  if the minimum distance between them satisfies the inequality $\delta k_F a >|\Delta|/E_F$.) The corresponding expressions for the bulk contribution to $J_2$ are
\begin{align}
     \bulk[2] =\bulkunit[2]\times \frac {t_\perp^4} {|\Delta| E_F^2} \times \left(\frac {|\Delta|} {E_F}\right)^{3\delta}
     \label{eq:bulktwo-scale}
\end{align}
In contrast, because (one component of the) momentum is not conserved for tunnelling at the edge (see Eq. \eqref{eq:edge-contr} and Fig. \ref{fig:FS}c,d,  \ref{fig:rotation-sim}a,b), the magnitude of the edge contribution does not depend as sensitively on the relative positions of the Fermi surfaces, so generally 
\begin{align}
    \edge =\edgeunit\times \frac{t_\perp^2 |\Delta|}{E_F^2}.
    \label{eq:edge-scale}
\end{align}


Ideally, measurements of the Josephson effect should permit separate evaluation of $J_1$ and $J_2$ \cite{golubov2004current};  in this case, when the bulk contribution to $J_1$ vanishes by symmetry, the symmetry sensitive angle dependence of $\edge$ can be used as a phase-sensitive probe of the order parameter of the lower SC.  
However, even if this is not possible, there exists a range of circumstances in which the edge contribution dominates the total Josephson energy, provided $t_\perp$ is small enough, and if the junction area is not too large.  Specifically, so long as $\bulk$ vanishes,
\begin{equation}
    \frac{J_1}{J_2}  = \frac{\edgeunit}{\bulkunit[2]} \left(\frac{|\Delta|}{t_\perp} \right)^2 \left( \frac{|\Delta|}{E_F}\right)^{3\delta} \times \frac{La}{A}.
\end{equation}
In other words, the edge contribution dominates the higher order bulk contribution so long as
\begin{equation}
    \label{eq:constraint-bulktwo}
    \frac{A}{La} \ll \left(\frac{|\Delta|}{t_\perp} \right)^2  \left( \frac{E_F}{|\Delta|}\right)^{3\delta}
\end{equation}
Note that this condition is much less stringent if the the FSs do not intersect ($\delta=1$) than if they do ($\delta=0$).  

\subsection{Disorder}
Since tunneling matrix elements are exponentially sensitive to local details, we consider random position-dependent variations in the tunnelling matrix element, $t_\perp(\br)$, to be the most important form of disorder.
\revise{(We have not explicitly considered another possibly important type of disorder, involving  variations in the gaps, $\Delta^\pm$, espeically if correlated between the two laters.) }
As discussed in greater detail in Supplementary Note 1, we 
consider the situation in which there is a mean tunneling matrix, $t_\perp \equiv \overline{t_\perp(\br)}$, and a variance, $\overline{[t_\perp(\br)-t_\perp][t_\perp(\br')-t_\perp]}$ 
that is short-range correlated with mean squared value \revise{$g_i^2$ and range $\xi_i$, where $i=\text{b},\text{e}$} depending on whether we are considering the bulk or edge regions.
When computing the disordered-average first order Josephson coupling  \footnote{It is not necessary to assume Gaussian distribution since $\bar{J}_1$ is due to tunneling of a single Cooper pair and thus does not involve higher order correlations}, it is straightforward to see that that the clean-limit and disorder contributions add, i.e.,
\begin{equation}
    \bar{J}_1 = (\bulk +\delta\bulk) \frac{A}{a} + (\edge +\delta \edge) \frac{L}{a}+\cdots
\end{equation}
where $\bulk,\edge$ are given as before.

Most importantly, the symmetry considerations that control $j_{1i}$ apply to $\delta j_{1i}$ as well  \footnote{\revise{In fact, the symmetry considerations for the disordered system can be proven in a more general setting, in which only time-reversal symmetry is assumed (so that multi-band non-local tunneling matrices can be included). See Supplementary Note 1 for details.}} since, although any given disorder realization locally breaks the point-group symmetries, the ensemble is assumed to have the same symmetries as the crystal, and so therefore do averaged quantities.
Even the  scaling of the quantities, 
i.e. when not forbidden by symmetry, 
$\delta j_{1i}\sim \left(g^2_i|\Delta| / E_F^2\right)\left( |\Delta| / 
E_F\right)^{\delta_i}$,
is typically not greatly affected by disorder so long as the correlation length is substantial, i.e. $k_F\xi \gg 1$.  
The essential effect of disorder is to give a momentum boost of order $\delta k \sim 1/\xi$.  
Thus $ \delta_i$ is determined in the same way as $\delta$ where if the distance between the important portions of the Fermi surfaces are spaced by $\delta k_F > \delta k$, these portions can be treated as non-intersecting, while otherwise they act as though they are intersecting.  

\subsection{Microscopic model results}
The geometry of the problem we have in mind, shown in Fig. \ref{fig:setup}a, consists of  a full-plane 2D layer superconductor,
on top of which, we place a half-plane superconductor so that there is an edge along $x=0$.
We will always compute the Josephson energy perturbatively in powers of the tunnelling Hamiltonian that couples the two layers. 
However, to make the calculations simpler, we will consider a simplified model consisting of two full-plane superconductors, but with a position dependent tunneling Hamiltonian that is  nonzero only in the half-plane $x >0$ (see Fig. \ref{fig:setup}b).
The latter setup means that the unperturbed Hamiltonian is translationaly invariant so that momentum $\bk$ is a good quantum number.

The Hamiltonian thus has the form
\begin{equation}
    \sH = \sH_0 + \sH_T
\end{equation}
where $\sH_T$ is the interlayer tunneling Hamiltonian, 
assumed to be spatially local and to have  
 magnitude $t_\perp$ for $x>0$ and vanish for $x<0$. Here
$\sH_0 = \sH_0^+ +\sH_0^-$ describes the decoupled bilayer superconductors with the top/bottom $\sH_0^\pm$ characterized by the properties $\bg^\pm = (\ve^\pm, \Delta^\pm)$ where $\ve^\pm (\bk)$ are the normal state dispersions and $\Delta^\pm (\bk)$ are the gap functions \footnote{In general, each SC gap $\Delta^\pm$ can possess a $U(1)$ phase and thus $\alpha$ denotes the phase difference. However, since the coefficients $J_1,J_2$ do not depend on $\alpha$, we have refrained from introducing more notation.}.
For analytical formulas, we assume that the junction is SBSP, while in all numerical calculations, we considere an SBSP $s$-$d_{x^2-y^2}$ junction with $E_F=1$ and 
\begin{align}
    &\ve^+(\bk) = -S_1(\bk)-S_2(\bk)-\mu^+,\nonumber\\
    &\ve^-(\bk) = -S_1(\bk) - \mu^- ,\\
    &\Delta^+(\bk) = \Delta_s, \ \ \Delta^-(\bk)=\Delta_d D(\bk),\nonumber
\end{align}
(The general case is treated in Supplementary Note 1).
In terms of the  lattice harmonics, $S_1(\bk) = \cos(k_x)+\cos(k_y)$, $S_2(\bk)= 2\cos(k_x)\cos(k_y)$, and $D(\bk) = \cos(k_x)-\cos(k_y)$.
The values of $\mu^\pm$ used are listed in each figure where numerical results are presented;  in numerically evaluating the integral  expressions for $\edge$ and $\bulk$ we have typically taken suitably small (compared to the bandwidth) values of $\Delta_s=0.12$ and $\Delta_d=0.05$.

\subsubsection{Ideal alignment $\theta=\phi=0$ and Scaling}
\label{sec:ideal-alignment}

To introduce the problem let us first consider a  junction in which the crystalline axes are aligned $\theta=0$ and the edge is also aligned $\phi=0$ 
(see Fig. \ref{fig:angles}).
\revise{For SBSP SCs, to second order in perturbation theory with respect to $t_\perp$ (see Supplementary Note 1),
\begin{equation}
    \label{eq:bulk-contr}
    \bulk =  t_\perp^2 j(\bq=\bm{0} )
\end{equation}
Where
\begin{align}
    j(\bq) &=  \iint_{\dB\dZ} f(\bk,\bq) \frac{d^2\bk}{(2\pi)^2} \\
    f(\bk,\bq) &= \frac{1}{E^+(\bk -\bq) +E^-(\bk)} 
    \frac{\Delta^+(\bk-\bq)}{E^+(\bk-\bq)} \frac{\Delta^-(\bk)}{E^-(\bk)} \nonumber
\end{align}
With $E^\pm(\bk) = \sqrt{\ve^\pm(\bk)^2+|\Delta(\bk)|^2}$.
Physically, the process that contributes to $\bulk$ involves an intermediate excited state in which one quasi-particle of momentum $\bk$ is created in one plane, and another with momentum $\bk$ in the other;  $\bulk$ is thus obtained by integrating these processes over all $\bk$.  
The energy of the excited state is  $E^+(\bk) + E^-(\bk )$ so the integral tends to be dominated by points in $\bk$ space at which this energy is smallest.  
For physically reasonable conditions in which $|\Delta|\ll E_F$, this means points in $\bk$ space that lie on both Fermi surfaces, if such points exist.  The actual integral - including the explicit dependence on the order parameter symmetry - comes through matrix elements related to the usual coherence factors of BCS theory.    
The calculation of $\bulk[2]$ proceeds similarly, but with intermediate states that involve twice the number of quasi-particles, and correspondingly a double integral over $\bk$ and $\bk^\prime$.
}

\revise{
The structure of the expression for $\edge$ is similar to that for $\bulk$, but in this case the edge can act as a source of momentum non-conservation, i.e., (see Supplementary Note 1)
\begin{equation}
    \label{eq:edge-contr}
    \edge = 2 t_\perp^2 \int_{-\pi}^\pi \frac{\delta \sj(q\hat{\underline{\bx}})}{q^2} \frac{dq}{2\pi}  -\frac{2}{\pi^2} \bulk  
\end{equation}
Where $\hat{\underline{\bx}}$ is the reciprocal primitive vector perpendicular to the edge and
\begin{align}
    \label{eq:dj}
    \delta \sj(\bq) &=  \left[\frac{|\bq|/2}{\sin(|\bq|/2)} \right]^2\frac{j(\bq)+j(-\bq)}{2} -j({\bf 0})
\end{align}
Consequently, the intermediate states consist of one quasi-particle with momentum $\bk$ in one SC and one with momentum $\bk-\bq$ in the other, where $\bq$ is a vector perpendicular to the interface.  
The full result for $\edge$ thus involves both a two dimensional integral over $\bk$ and a one-dimensional integral over $\bq$.  
In this case, the energy of the excited state is $E^+(\bk -\bq) + E^-(\bk)$, so the integral is dominated by points at which $\bk$ lies on the Fermi surface of one SC and $\bk-\bq$ on the Fermi surface of the other.
Also note that $\edge$ in Eq. \eqref{eq:edge-contr} includes $\bulk$, but since we will be concerned with scenarios where the bulk $\bulk =0$ (e.g., $s$-$d$ wave junction), it will not affect the overall scaling.
}

\begin{figure}[ht]
\centering
\includegraphics[width=1\columnwidth]{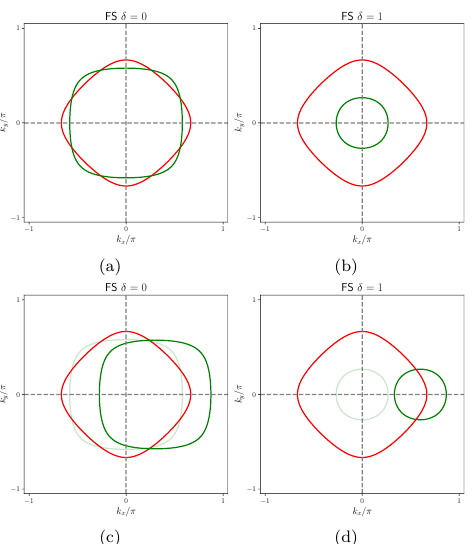}
\caption{Fermi surface geometries for $\theta=\phi=0$. 
The red and green contours represent, respectively the top and bottom layer FSs, where (a) illustrates a case in which there are points of intersection and (b) where there are none.
In (c) and (d), the top layer FSs from (a) and (b) are  shifted by momentum $\bq$ perpendicular to the edge. 
Here, $\mu^- = -0.5$ for all subplots, while $\mu^+ = -0.3$ for (a),(c) and $\mu^+=-3$ for (b),(d).
}
\label{fig:FS}
\end{figure}

As discussed following Eqs. \eqref{eq:bulk-scale},\eqref{eq:bulktwo-scale}, the bulk contributions $\bulk$, $\bulk[2]$, etc. scale differently depending on whether the FS of the top and bottom layers intersect (as shown schematically in Fig. \ref{fig:FS}a) or not (as in Fig. \ref{fig:FS}b).  
In the former case, the integrals of $\bk$ are dominated by regions in the neighborhood of the intersections, where the energy denominators are of order $|\Delta|$ (and the coherence factors, as well, tend to be relatively large).
In the latter case, all energy denominators are of order $E_F$, leading to an additional power of $|\Delta|/E_F$ in the parametric dependence  of these quantities, i.e. the exponent $\delta$ is $\delta=0$ in the first case and  $\delta=1$ in the latter.

In contrast,  in computing $\edge$ one necessarily averages over relative momentum shifts $\bq$ of the FS, as shown in Figs. \ref{fig:FS}c,d. 
This typically leads to FS crossings, even when the unshifted FSs are non-overlapping, and is responsible for the fact that the parametric dependences of $\edge$ are relatively insensitive to details of the FS structure.
\revise{
To emphasize this relation, we write \footnote{Recall, that if a symmetry operator $\sS$ (say, translation or rotation) acts on a space (e.g., $\bk \in \dB\dZ$), then it induces a symmetry operation on functions over the space (e.g., $\ve^\pm (\bk)$) via the relation $(\sS \ve^\pm)(\bk) \equiv \ve^\pm (\sS^{-1}\bk)$. Note $\sS^{-1}$ is invoked so that the dirac-delta function $|\bk\ket$ centered at momentum $\bk$ is mapped via $\sS|\bk\ket =|\sS\bk\ket$.} $j(\bq) = j[\tau_{\bq} \bg^+,\bg^-]$ where $\tau_{\bq}$ denotes the translation operator $\bk \mapsto \bk +\bq$ in momentum space and $\bg^\pm =(\ve^\pm, \Delta^\pm)$ characterize the upper/lower SCs.
}

\subsubsection{General case $\theta, \phi \neq0$ and Symmetry}
\label{sec:general-rotation}

\begin{figure}[ht]
\centering
\includegraphics[width=0.7\columnwidth]{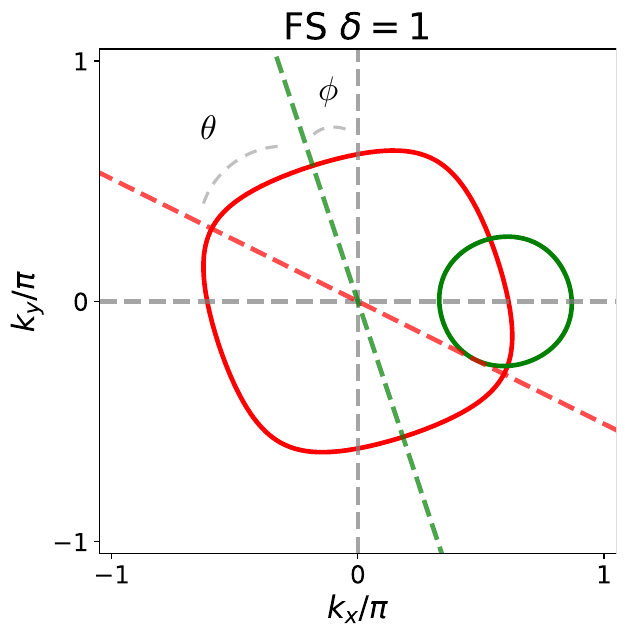}
\caption{\revise{Generic junction geometry. 
  In a coordinate system aligned with the edge in the top SC, the green curve indicates the upper layer FS, which is first rotated by angle $\phi$, then shifted by $\bq$ perpendicular to the edge, while the red curve indicates the bottom layer FS, which is rotated by angle $\theta+\phi$. 
  The dashed green and red lines indicate the rotated principle axes $\hat{\by}$ of the upper/lower SCs.}
}
\label{fig:rotation-new}
\end{figure}

\revise{
To extend the expressions of $\bulk,\edge$ in Eq. \eqref{eq:bulk-contr}, \eqref{eq:edge-contr} to general $\theta,\phi$, the \textit{morally} \footnote{At general angles, there is no underlying Bravais lattice that characterizes both SC layers simultaneously and thus, in principle, the previous expression of $\bulk,\edge$ in Eq. \eqref{eq:bulk-contr}, \eqref{eq:edge-contr} are not exactly correct.
However, since the integrals are dominated by momentum near the FS, this technicality will be neglected.} 
correct approach is to rotate the upper/lower SCs $\bg^\pm \mapsto \sR_\phi \bg^+, \sR_{\theta+\phi} \bg^-$ in $j(\bq)$ where $\sR_\phi$ denotes rotation by $\phi$ as illustrated in Fig. \ref{fig:rotation-new}.
For example, by Eq. \eqref{eq:bulk-contr}, 
\begin{equation}
    \bulk(\theta) =j[\sR_\phi \bg^+, \sR_{\theta+\phi} \bg^-]=j[\bg^+,\sR_\theta \bg^-]
\end{equation}
Where equality follows from the integral being invariant under simultaneous rotation in $\bg^\pm$.
In particular, one can derive the symmetry consideration
\begin{align}
    j[\bg^+,\sR_{\theta} \bg^-] &= j[C_4\bg^+, C_4 \sR_\theta \bg^-] \\
    &=  j[C_4 \bg^+,\sR_\theta C_4 \bg^-]\\
    \bulk(\theta) &= -\bulk(\theta) \label{eq:bulk-vanishes}
\end{align}
And thus $\bulk(\theta)=0$ for all twist angles. 
In contrast, the edge breaks $C_4$ symmetry and thus $\edge$ is generally nonzero.
The remaining symmetry considerations discussed in the introduction can be similarly derived (see Supplementary Note 1 for details).
}

\begin{figure}[ht]
\centering
\includegraphics[width=1\columnwidth]{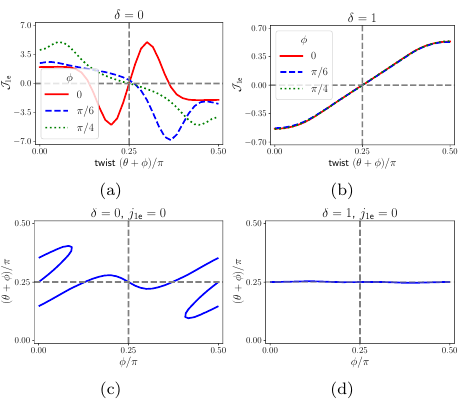}
\caption{Results computed for the model band structure.  For subplots (a),(c), the normal state FSs are assumed to be as in Fig. \ref{fig:FS}a at ideal alignment $\theta=\phi=0$, while for subplots (b),(d), the normal state FSs are those in Fig. \ref{fig:FS}b at ideal alignment.
(a),(b) plots the dimesionless magnitude of the edge contribution $\edgeunit$ as a function of edge angle $\theta +\phi$ (relative to the lower layer) for distinct edge angles $\phi=0,\pi/6,\pi/4$ (relative to upper layer). 
(c),(d) plots the critical angles at which the edge vanishes $\edge=0$ with the $y$ axis being $(\theta+\phi)/\pi$.}
\label{fig:rotation-sim}
\end{figure}

\revise{
The parametric scaling of $\bulk,\bulk[2],\edge$ at generic values of $\theta$ and $\phi$ follows the same argument as that at ideal alignment $\theta=\phi=0$, provided that it is not forbidden by symmetry.
In particular, numerical simulations shown in Fig. \ref{fig:rotation-sim}a,b illustrate that $\edge$ is generically nonzero and satisfy the scaling in Eq. \eqref{eq:edge-scale}.
Moreover, since the upper layer is assumed to be a ($D_4$ symmetric) $s$-wave, in the limiting case where the top layer FS is a singular point, the top SC obtains full $O(2)$ symmetry and thus the $\edge (\theta,\phi)$ only depends on the edge angle $\theta+\phi$ relative to the lower layer, i.e., $\edge (\theta,\phi) = \edge(\theta+\phi)$.
Away from the fine-tuned limit, numerical calculations in Fig. \ref{fig:rotation-sim}b confirm that this approximation, $\edge(\theta,\phi) \approx \edge (\theta+\phi)$, is valid for small but realistically sized FSs.
In particular, Fig. \ref{fig:rotation-sim}d shows that $\edge$ vanishes near $\theta +\phi \approx \pi/4 \mod \pi/2$.
For a larger FS, the approximation breaks down as shown in Fig. \ref{fig:rotation-sim}a,c.
}

\subsubsection{Other  Considerations}

\begin{figure}[ht]
\centering
\includegraphics[width=1\columnwidth]{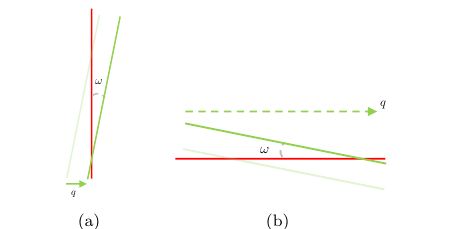}
\caption{Small angle $\omega$ intersection of FS.}
\label{fig:small-angle}
\end{figure}
There are a variety of more detailed features of the electronic structure that effect the magnitude of 
$\bulk$, $\bulk[2]$ and 
$\edge$: 

\begin{enumerate}[label=\textbf{\arabic*)}]
\item  \textbf{Number of FS intersections}:
As already pointed out, there is a parametric difference between the case of no FS crossings (for which $\delta=1$) and any finite number (for which $\delta=0$).
When computing the edge contribution $\edge$, the top and bottom FSs can intersect at general momentum shifts $\bq$.
In computing $\edge$, if the unshifted ($\bq=0$) FSs are non-intersecting, the shifted FSs will typically intersect at 2 points as shown in Fig. \ref{fig:FS}c,d.
Conversely, when the FSs are intersecting $\delta=0$, by $C_4$ symmetry, the two FSs generally intersect at $8=4\times 2$ points (2 in each quadrant) and thus there exists a factor $\times 8$ enhancement in both the bulk and edge contributions (in all orders of perturbation $\bulk,\edge,\bulk[2],...$).

\item \textbf{Angle of intersection}: 
The magnitude of the couplings also depends on the angle $\omega$ between the two intersecting pieces of the FSs.
The ``nested'' case in which $\omega=0$,  would lead to a parametrically large (in $E_F/|\Delta|$) enhancement.
However, since this is a  fine-tuned occurrence, we will consider $\pi\gg \omega>0$, and look at the scaling as $\omega \to 0^+$. It is straightforward to verify that the bulk coefficients are all enhanced by a factor of $ |\pi/\omega|$.
However, the edge coefficient $\edge$ may or may not gain an extra factor depending on the direction of intersection.
For example, in the schematic sketch shown in Fig. \ref{fig:small-angle}a, $\edge$ is not so enhanced, since the factor $1/\omega$ associated with the particular value of $\bk$ for which the nesting condition is satisfied, is offset by a factor of
$\omega$ from the range of $\bq$ over which this condition is approximately satisfied. 
Conversely, if the two FS were to intersect horizontally at small angles as sketched in Fig. \ref{fig:small-angle}b, then the edge coefficient $\edge$ gains an extra factor $1/\omega$ enhancement (not $1/\omega^2$ since the momentum regime at which the two FS intersect is bounded above, i.e., $\Delta q \le 2\pi$).
\end{enumerate}
\section{Discussion}

\begin{figure}[ht]
\centering
\includegraphics[width=0.6\columnwidth]{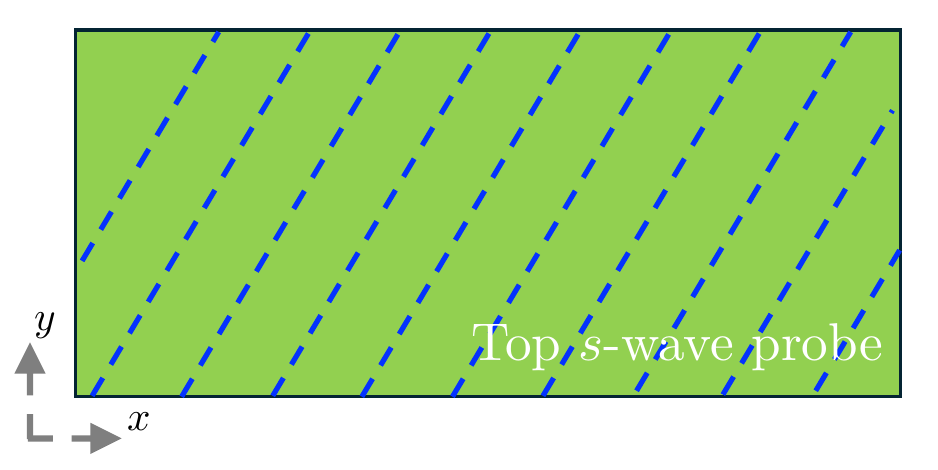}
\caption{Schematic. The parallel dashed lines resemble the artificially created edges (possibly through laser damage) in the top layer probe (in comparison to Fig. \ref{fig:setup}).
}
\label{fig:real-setup}
\end{figure}

In a realistic setup, the upper layer probe is a finite plane with multiple edges at distinct edge angles rather than a single edge as described by the half-plane in Fig. \ref{fig:setup}a.
In this case, the edge contribution $\edge$ is the sum of that for each edge of the finite probe and thus an ideal experimental setup would involve a probe geometry forming a long parallelogram so that the edge contribution is mainly due to the pair of long edges (and thus a single edge angle $\phi$ is well-defined).
Another possibly way \footnote{An idea proposed by Philip Moll in private conversation.} to construct such junctions with control over the geometric factors $\phi$ and $A/L$ is to use lasers to damage a large planar probe along parallel lines (as sketched in Fig. \ref{fig:real-setup}), thereby creating artificially oriented edges within the junction.

\revise{
As discussed in the introduction, in principle, any SC probe with known symmetry can be used to reveal the lower SC symmetry, regardless of the presence of disorder.
However, we note that the ideal probe should be an $s$-wave SC with a small FS. 
Indeed, the lowest order edge contribution $\edge$ dominates the higher order bulk $\bulk[2],...$ provided that the constraint in Eq. \eqref{eq:constraint-bulktwo} is satisfied and thus if the distance $\delta k_F$ between the two FS is large, the constraint is much easier to satisfy.
Furthermore, in the fine-tuned scenario where the FS consists of a single point, a non-$s$-wave SC symmetry would imply that the effective gap vanishes. For reasonably small FS, a non-$s$-wave probe implies a unfavorable decrease in magnitude of $\edge$.
}

There are a large number of circumstances in which unusual SC states are thought to arise in (quasi-) 2D systems, including twisted bilayer graphene \cite{cao2018correlated,cao2018unconventional,yankowitz2019tuning,andrei2020graphene} and its cousins \cite{zhao2023time,yuan2023inhomogeneity,yuan2023exactly,yuan2024absence}, for  which our proposed approach might prove useful in extracting many important features of the structure of the SC state \footnote{
In fact, since Moire systems have small FSs, one usually encounters difficulty in finding a SC probe with similar size FS so that FSs can intersect (and so the bulk contribution $\bulk$ is significant). In contrast, in the half-plane setup in Fig. \ref{fig:setup}, momentum is not conserved perpendicular to the edge and thus the scaling of the edge contribution $\edge$ is independent of the relative sizes of the FSs.}.
One possible application is to determine the order parameter symmetry of \sro{}, a longstanding enigma spanning over three decades \cite{maeno2024still}.
Indeed, previous proposals have floated the notion of $d+ig$ pairing symmetry \cite{kivelson2020proposal,yuan2021strain,willa2021inhomogeneous} where the $g$-wave predominantly occupies the $\alpha,\beta$ bands of \sro{} \cite{yuan2023multiband}.
Since the $\alpha,\beta$ bands comprises $d_{xz},d_{yz}$ orbitals, the corresponding $c$-axis tunneling parameter $t_\perp$ is much larger than that of the $\gamma$ band which consists of $d_{xy}$ orbitals.
Consequently, it is possible that the $c$-axis Josephson junction could serve as a means to scrutinize the dominant pairing symmetry on the $\alpha,\beta$ bands.

Another interesting application is to probe the magnitude of the SC gap along the FS, since the magnitude of the coupling depends rather sensitively on details of the FS structure.
For instance, consider the case in which the SC of interest (the lower SC) is trivial ($s$-wave) in terms of symmetry, but is thought to have significant angle-dependence of $\Delta$. 
Such examples occur in (slightly) orthorombic materials where $s$-wave is the only irrep of the group symmetry, but since the material is almost tetragonal, it is conceivable that the SC gap inherits a $d+s$ wave symmetry from the tetragonal group symmetry.
We could explore this by choosing the probe (upper) SC to have two Fermi pockets surrounding points $\pm \bm{\mathrm{Q}}$ (so that the Fermi pockets intersect the FS of the lower SC);  then as a function of $\theta$, the portion of the FS of the lower SC that principally dominates $\bulk$ would change and the angle-dependence of $\Delta^-(\bk)$ would be thus revealed.  
Similarly, by varying $\phi$, one can extract similar information from $\edge$ (though in this case, the Fermi pockets should be away from the FS of the lower SC so that the bulk $\bulk$ contribution is suppressed as in Eq. \eqref{eq:bulk-scale}).

\section{Methods}
All technical aspects of our analysis are coherently presented in the ``Results'' and ``Discussion'' sections.

\section{Data Availability}
The data that supports the findings of this study are available from the corresponding author upon reasonable request.

\section{Acknowledgements}

We acknowledge extremely helpful discussions with Philip Kim, Daryl Schlom, Brad Ramshaw, Kathryn  Moler, Phillip Moll, Malcolm  Beasley, and Sankar Das Sarma.
A.C.Y. and S.A.K. were supported, in part,  by NSF grant No.  DMR-2310312 at Stanford. 

\section{Author Contributions}
A.C.Y and S.A.K inititated the project and conceived the presented idea. A.C.Y. carried out the calculations. A.C.Y. and S.A.K. wrote the paper. The manuscript reflect contributions from all authors.

\section{Competing Interests}
The authors declare no competing interests.

\bibliography{main.bbl}

\end{document}


\title{Supplemental information for Phase sensitive information from a planar Josephson junction}
\author{Andrew C. Yuan}
\affiliation{Department of Physics, Stanford University, Stanford, CA 94305, USA}
\author{Steven A. Kivelson}
\affiliation{Department of Physics, Stanford University, Stanford, CA 94305, USA}

\maketitle
\tableofcontents


\begin{center}
\textbf{Supplementary Note 1}
\end{center}
\section{Single Band Singlet Pairing (SBSP)}

In this section, we shall consider the simplifed case where we not only assume time-reversal symmetry (TRS), but also assume that the superconductors (SC) are single-band with singlet pairing. The general case is postponed till Sec. \eqref{app:general}.

\subsection{Computing $\bulk,\edge$}
\label{app:edge-comp}

\begin{figure}[ht]
\subfloat[\label{fig:physical-setup}]{%
\centering
\includegraphics[width=0.48\columnwidth]{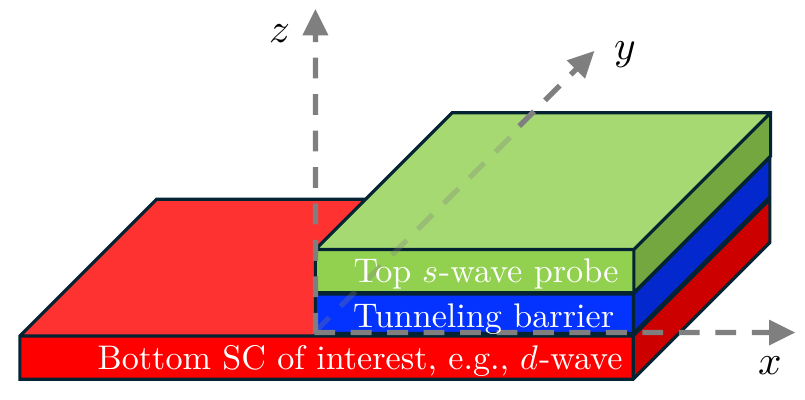}
}
\subfloat[\label{fig:equi-setup}]{%
\centering
\includegraphics[width=0.48\columnwidth]{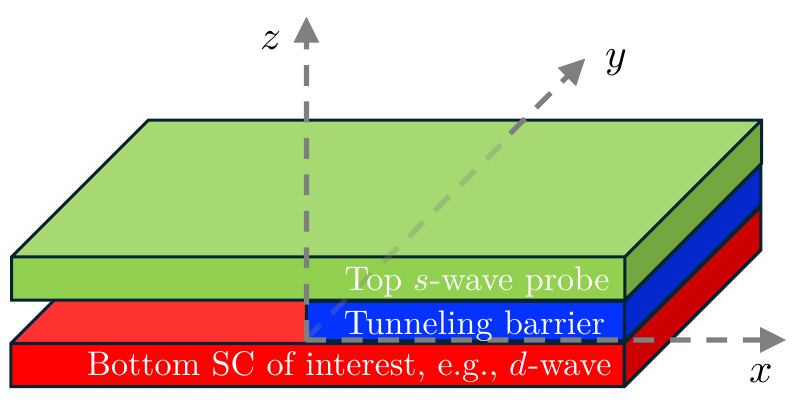}
}
\caption{Junction geometry.  (a) shows the physical geometry where the upper SC terminates at an edge at $x=0$.  (b) shows the model geometry used to simplify microscopic calculations, in which both SCs are infinite, but tunnelling between them is only permitted for $x>0$.
}
\label{fig:setup}
\end{figure}

Using the simplified setup shown in Fig. \ref{fig:equi-setup} (first presented in the main text but repeated here for convenience), at ideal alignment $\theta =\phi=0$, we take each layer to have linear size $L=2\ell$ (with even $\ell$ for simplicity) and adopt periodic boundary conditions so that translation symmetry is preserved.
The general formula for the first order Josephson coupling $J_1$ for single band singlet SCs \revise{(see Eq. \eqref{app-eq:J1-general-k})} is then given by
\begin{equation}
    \label{app-eq:J1-singlet}
    J_1 = \sum_{\bk',\bk\in \dB\dZ} |T(\bk',\bk)|^2 f(\bk,\bk-\bk')
\end{equation}
where $f(\bk,\bq)$ is (given in the main text, but repeated here for convenience) 
\begin{equation}
    f(\bk,\bq) = \frac{1}{E^+(\bk -\bq) +E^-(\bk)} 
    \frac{\Delta^+(\bk-\bq)}{E^+(\bk-\bq)} \frac{\Delta^-(\bk)}{E^-(\bk)}
\end{equation}
and $T(\bk',\bk)$ is the interlayer tunneling matrix in momentum space, i.e., the tunneling matrix is diagonal in real space \revise{$\br=x\hat{\bx} +y\hat{\by}$ (where $\hat{\bx},\hat{\by}$ are the primitive vectors of the underlying Bravais lattice with reciprocal primitive vector $\hat{\underline{\bx}},\hat{\underbar{\by}}$)} and $=t_\perp$ only on the half-plane $x\ge 0$, and thus
\begin{align}
    \frac{T(\bk',\bk)}{t_\perp} &=\frac{1}{L^2} \sum_{x\ge 0} e^{-i(\bk'-\bk)\br} \\
    &= \frac{1}{L} 1\{\bk -\bk'=q\hat{\underline{\bx}}\} \sum_{x =0}^{\ell-1} e^{iqx} \\
    &= \frac{1}{2} 1\{\bk'=\bk\} + 1\{\bk-\bk'= q\hat{\underline{\bx}}, q\; \text{odd}\} \frac{2}{L} \frac{1}{1-e^{iq}} \nonumber
\end{align}
where \revise{$1\{A\}$ denotes the indicator function so that it $=1$ if condition $A$ is satisfied and $=0$ otherwise, while} the momentum transfer during tunelling $\{q\;\text{odd}\}$ is short for $q=2\pi m/L$ where $m$ is an odd integer in $\{-\ell, ...,\ell-1\}$.
Therefore,
\begin{align}
    \label{app-eq:J1}
    J_1 &= \underbrace{\frac{t_\perp^2}{4}\sum_{\bk\in \dB\dZ} f(\bk,\bq=\bm{0})}_{(1)} + \underbrace{t_\perp^2 \sum_{q\;\text{odd},\bk \in \dB\dZ} \frac{1}{L^2} \frac{1}{\sin^2 (q/2)} f(\bk,q\hat{\underline{\bx}})}_{(2)}.
\end{align}
Note that
\begin{equation}
    \label{app-eq:j-def}
    \frac{1}{L^2} \sum_{\bk \in \dB\dZ} f(\bk,q\hat{\underline{\bx}}) = \underbrace{\iint_{\dB\dZ} f(\bk,q\hat{\underline{\bx}})}_{j(q\hat{\underline{\bx}})} +O\left(\frac{1}{L^2}\right).
\end{equation}
Where we have kept the error term since we are interested in not only the leading order term $\sim L^2$ but also the next to leading term, $\sim L$.
Hence, the first term in Eq. \eqref{app-eq:J1} contributes a bulk effect, i.e.,
\begin{equation}
    (1) = \frac{t_\perp^2}{4} j(\bm{0}) L^2 +O(1).
\end{equation}
The second term is given by
\begin{align}
    (2) &= t_\perp^2 \sum_{q\;\text{odd}} \frac{1}{\sin^2 (q/2)} \left[j(q\hat{\underline{\bx}}) +O\left(\frac{1}{L^2}\right)\right].
\end{align}
It turns out that this 
also contains a $(L^2)$ bulk contribution  in addition to an edge $(L)$ contribution.
To see this, focus on the small $q$ portion of the sum, and observe that
\begin{align}
    \label{app-eq:q2-div}
    \sum_{q\;\text{odd}} \frac{1}{q^2} = \frac{L^2}{2\pi^2}  \underbrace{\sum_{m=0}^{\ell/2-1} \frac{1}{(2m+1)^2}}_{\le \pi^2/8}.
\end{align}
Since the summation is bounded by $\pi^2/8$ in the $L\to \infty$ limit, we see that the $1/q^2$ divergence is at most $O(L^2)$ and thus we can drop the error term since it is now $O(1)$, i.e.,
\begin{align}
    \label{app-eq:edge-term}
    (2) &= t_\perp^2 \sum_{q\;\text{odd}} \frac{j(q\hat{\underline{\bx}})}{\sin^2 (q/2)}   +O(1)
\end{align}
To treat the first term in Eq. \eqref{app-eq:edge-term}, write
\begin{align}
    \sj(\bq)  &= \left( \frac{|\bq|/2}{\sin (|\bq|/2)} \right)^2 j(\bq)\\
    \delta \sj(\bq) &= \frac{1}{2} (\sj(\bq)+\sj(-\bq))-\sj(\bm{0})
\end{align}
Where we note that $\sj(\bm{0})=j(\bm{0})$. 
Therefore,
\begin{align}
    \label{app-eq:edge-term-simp}
    \sum_{q\;\text{odd}} \frac{j(q\hat{\underline{\bx}})}{\sin^2 (q/2)}   &=  j (\bm{0})  \sum_{q\;\text{odd}} \frac{4}{q^2} + \sum_{q\;\text{odd}} \frac{4}{q^2} \delta \sj(q\hat{\underline{\bx}}) 
\end{align}
Note that the second term is now well-regulated in the thermodynamic limit $L\to \infty$ since $\delta \sj(q\hat{\underline{\bx}})\propto q^2$ for small $q$, and thus
\begin{equation}
    \frac{1}{L} \sum_{q\;\text{odd}} \frac{4}{q^2} \delta \sj(q\hat{\underline{\bx}}) = 2\int_{-\pi}^\pi \frac{\delta \sj(q\hat{\underline{\bx}})}{q^2} \frac{dq}{2\pi} +O\left(\frac{1}{L} \right).
\end{equation}
The first term in Eq. \eqref{app-eq:edge-term-simp} 
is $O(L^2)$ 
and thus makes another (equal) contribution to the bulk effect as expected. 
To obtain the edge effect, we must consider higher orders of $1/L$ in the summation
\begin{align}
    \sum_{m=0}^{\ell/2-1} \frac{1}{(2m+1)^2} &= \left[\sum_{m=0}^{\infty} -\sum_{m\ge \ell/2}\right]\frac{1}{(2m+1)^2} \\
    &= \frac{\pi^2}{8} -\sum_{m\ge \ell/2}\frac{1}{(2m+1)^2}
\end{align}
Note that the 2\ts{nd} term scales like
\begin{equation}
    \int_{s > \ell } \frac{ds}{s^2} \sim \frac{1}{L}
\end{equation}
Therefore, to obtain the edge effect, we write
\begin{align}
    \sum_{m\ge \ell/2}\frac{1}{(2m+1)^2} &= \sum_{m\ge \ell/2} \frac{1}{(2m+1)(2m-1)}  +O\left(\frac{1}{L^2}\right) 
\end{align}
Where the difference is $O(1/L^2)$ since it scales like
\begin{equation}
    \int_{s >\ell} \frac{ds}{s^3} \sim \frac{1}{L^2}
\end{equation}
It is then straightforward to verify that
\begin{align}
    \sum_{m\ge \ell/2} \frac{1}{(2m+1)(2m-1)} &= \frac{1}{2}\sum_{m\ge \ell/2} \left(\frac{1}{2m-1}-\frac{1}{2m+1}\right) \\
    &= \frac{1}{2} \frac{1}{\ell-1}
\end{align}
Putting everything together, we find that in the limit $L\to \infty$, 
\begin{align}
    \sum_{q\;\text{odd}} \frac{4}{q^2} = \frac{1}{4} L^2 -\frac{2}{\pi^2} L +O(1)
\end{align}
Hence,
\begin{align}
    (2)  &=  \frac{t_\perp^2}{4} j(\bq = \bm{0}) \times L^2\\
    &\quad+
    t_\perp^2 \left[2 \int_{-\pi}^\pi \frac{\delta \sj(q\hat{\underline{\bx}})}{q^2} \frac{dq}{2\pi}  -\frac{2}{\pi^2} j(\bm{0})\right]\times L \\
    &\quad+O(1).
\end{align}
This  is of the form as the expressions given in the text,
\begin{equation}
    J_1 = \bulk A +\edge L +O(1)
\end{equation}
with  $A=L^2/2$  the area of half-plane.

\subsection{Symmetries}
\label{app:symmetry}

\begin{table}[ht!]
\begin{subtable}{.49\columnwidth}
\centering
\begin{tabularx}{.8\columnwidth}{C C C}
\toprule
$\theta \mapsto C_{2n}\theta$ & $\phi\mapsto C_{2n}\phi$ & $\theta,\phi\mapsto\sM\theta,\sM\phi$\\
\midrule
$r^-$ & $r$ & $m$\\
\bottomrule
\end{tabularx}
\caption{Symmetries of $\edge$}
\end{subtable}
\begin{subtable}{.49\columnwidth}
\begin{tabularx}{.8\columnwidth}{C C C C}
\toprule
$(0,0)$ &$\displaystyle\left(\frac{\pi}{2n},0\right)$ & $\displaystyle\left(0,\frac{\pi}{2n}\right)$ & $\displaystyle\left(\frac{\pi}{2n},\frac{\pi}{2n}\right)$ \\ [1.5ex]
\midrule
$m$ & $m r^+$ & $m r^-$ & $m r$ \\
\bottomrule
\end{tabularx}
\caption{Zeros of $\edge$}
\end{subtable}
\caption{$D_{2n}$ point group. Let $C_{2n},\sM$ denote $2n$-fold rotation and mirror symmetry along the edge $\hat{\by} \mapsto -\hat{\by}$ so that the SC gaps $\Delta^\pm \mapsto r^\pm \Delta^\pm, m^\pm \Delta^\pm$ (where $r^\pm,m^\pm \in \{\pm 1\}$) under $C_{2n},\sM$, respectively.  Write $r=r^+r^-$ and $m=m^+m^-$. 
(For example, in a tetragonal $s$-$g$ junction, $r^\pm = 1$ since $s,g$-waves are even under $C_4$ rotation, while $m^\pm=\pm 1$ since an $s,g$-wave is even/odd under mirror symmetry $\sM$, respectively.)
(a). The bottom row shows the symmetry of $\edge$ under transform given by the top row, e.g., $\edge \mapsto r^- \edge$ under $\theta \mapsto C_{2n}\theta$ while $\phi$ is kept invariant. In particular, $\edge$ depends on $\theta,\phi$ only modulo $\pi/n$. (b). The top row $(\phi,\theta +\phi) \mod \pi/n$ shows the angles that are highly symmetric due to the symmetries in (a). In particular, the top row is a zero of $\edge$ if the bottom row $=-1$. See SI for proof.
}
\label{tab:edge}
\end{table}

\revise{
In the main text, we claimed that the bulk contribution $\bulk(\theta)=0$ for all twist angles $\theta$ if the junction is odd under discrete rotation, and that the edge $\edge$ satisfies certain the symmetry considerations (repeated in Table. \eqref{tab:edge} for convenience). 
Here we provide a short proof for the illustrative scenario of SBSP SCs. 
Using the coordinate system with the edge aligned along $x=0$ and $\bulk = t_\perp^2 j(\bq =0)$, we see that
\begin{align}
    j[\bg^+,\sR_{\theta} \bg^-] &= j[C_{2n}\bg^+, C_{2n} \sR_\theta \bg^-] \\
    &=  j[C_{2n} \bg^+,\sR_\theta C_{2n} \bg^-]\\
    \bulk(\theta) &= r \bulk(\theta), \quad r=r^+r^-
\end{align}
Therefore, if the junction is odd $r=-1$ under discrete $C_{2n}$ rotations, the bulk contribution $\bulk(\theta)=0$ vanishes for all twist angles $\theta$.

For the edge contribution, note that
\begin{align}
    j[\tau_{q\hat{\underline{\bx}}}\sR_\phi \bg^+,\sR_{\phi +C_{2n}\theta} \bg^-] &= j[\tau_{q\hat{\underline{\bx}}}\sR_\phi\bg^+, \sR_{\phi+\theta} C_{2n} \bg^-] \\
    &= r^-j[\tau_{q\hat{\underline{\bx}}}\sR_\phi\bg^+, \sR_{\phi+\theta} \bg^-]\\
    \edge(C_{2n} \theta, \phi) &= r^- \edge(\theta, \phi) 
\end{align}
And that
\begin{align}
    j[\tau_{q\hat{\underline{\bx}}}\sR_{C_{2n}\phi} \bg^+,\sR_{C_{2n}\phi +C_{2n}\theta} \bg^-] &= j[\tau_{q\hat{\underline{\bx}}}\sR_\phi C_{2n} \bg^+, \sR_{\phi+\theta} C_{2n} \bg^-] \\
    &= r j[\tau_{q\hat{\underline{\bx}}}\sR_\phi\bg^+, \sR_{\phi+\theta} \bg^-]\\
    \edge(C_{2n} \theta, C_{2n} \phi) &= r \edge(\theta, \phi) 
\end{align}
And that
\begin{align}
    j[\tau_{q\hat{\underline{\bx}}} \sR_{\sM\phi} \bg^+,\sR_{\sM(\theta+\phi)} \bg^-] &= j[\tau_{q\hat{\underline{\bx}}} \sM \sR_{\phi} \sM \bg^+,\sM \sR_{\theta+\phi} \sM\bg^-] \\
    &= m j[\sM \tau_{\sM q\hat{\underline{\bx}}} \sR_{\phi}  \bg^+,\sM \sR_{\theta+\phi} \bg^-] \\
    &= m j[\tau_{q\hat{\underline{\bx}}}  \sR_{\phi}  \bg^+,\sM \sR_{\theta+\phi} \bg^-] \\
    \edge (\sM \theta,\sM\phi) &= m \edge(\theta,\phi)
\end{align}
Where we used the fact that mirror symmetry $\sM$ keeps the edge invariant. 
}
\subsection{Disorder}
\label{app:disorder}

In each order of perturbation theory, 
what enters are the tunneling matrix elements $T(\bk,\bk^\prime)$ for transferring an electron from a Bloch state with crystal momentum $\bk$ in the upper layer to $\bk^\prime$ in the lower layer - and naturally $T(\bk,\bk^\prime)^*$ associated with the reversed process.  (In more general circumstances, there could be multiple band indices associated with the single particle states of the decoupled layers, which we \revise{refer to Appendix \eqref{app:general} for full generality}.)  

When we consider the problem with disorder, we assume that mesoscopic effects can be neglected, or in other words we compute the configuration average of the various contributions to the Josephson coupling, $\bar{J}_n$.  
For the leading order term, $\bar{J}_1$, this means the results depend on \revise{$\overline{|T(\bk',\bk)|^2}$}, while higher order terms $n\ge 2$ need knowledge of higher order moments of $T$.  
In particular, for finite $L$ so that $\bk$ is a discrete variable,
\begin{equation}
    \bar{J}_1 = \sum_{\bk',\bk\in \dB\dZ} \ \overline{|T(\bk',\bk)|^2} \ f(\bk,\bk-\bk')
\end{equation}
where $f(\bk,\bq)$, which depends only on the properties of the decoupled layers, is given 
in Eq. \eqref{eq:dj}.

For the purposes of explicit calculations, as in the clean limit, we will assume that the tunnelling is defined by a local matrix element, $t(\br)$, so that
\begin{align}
    T(\bk,\bk')&\equiv \frac{1}{L^2}\sum_{\br} t(\br) e^{-i(\bk-\bk')\br}.
\end{align}
To account for disorder in the tunnelling matrix elements, we assume a random distribution with mean $\overline{t(\br)}$ and variance:
\begin{equation}
    \overline{t(\br) t(\br')}\ - \ 
    \overline{t(\br)}\times \overline{t(\br')}=\sigma(\br,\br'),
\end{equation}
In computing the first order Josephson coupling, the precise form of the distribution of $t(\br)$'s is unimportant - only these  moments matter.

\begin{enumerate}[label=\textbf{\arabic*)}]
    \item As a reminder, a system without edges only possesses a bulk contribution and thus it is standard to take the mean $\overline{t(\br)} = t_\perp$ and variance $\sigma(\br,\br')=\bdis(\br-\br')$ so that
    \begin{align}
       \overline{|T(\bk',\bk)|^2}= 
       |t_\perp|^2\ \delta_{\bk,\bk'} + \frac{1}{L^2}\Bdis(\bk-\bk')
    \end{align}
    where the first term denotes the clean-limit contribution proportional to $|t_\perp|^2$, and $\Bdis$, which is the Fourier transform of $\bdis$, reflects the point-to-point variations in the tunnelling matrix elements, and so is characterized by a magnitude, $\Bdis({\bf 0})$, and a width in $\bk$-space $|\delta \bk|\sim 1/\xi$.
    \item In comparison, when an edge is present (as shown in Fig. \ref{fig:equi-setup}), we need to take into account the lack of (even average) translational symmetry and thus take
    \begin{equation}
        \overline{t(\br)} = t_\perp \Theta(x), \quad \sigma(\br,\br') = \edis(\br-\br')\Theta(X)
    \end{equation}
    where $\Theta$ is the Heaviside function and $X=(x+x')/2$ is the average displacement from the edge \footnote{There is a subtlety in defining $\Theta(X)$ since $X\mapsto X+L/2$ under periodic translation $x\mapsto x+L$. This subtlety introduces an innate UV cutoff which results in an edge term contribution from the disorder. We discuss this in more detail in Appendix \eqref{app:edge-disorder-correlation}.}. 
    In particular, if $\br$ and $\br'$ are far from the edge, the disorder correlation reduces to a bulk term $\sigma(\br,\br') = \edis(\br-\br')$ (For example, it is reasonable to imagine $\sigma(\br,\br')=\bdis(\br-\br')\Theta(X)$.) \footnote{Although the current form of disorder $\sigma(\br,\br')$ gives both a bulk $\delta \bulk$ and edge $\delta \edge$ contribution, it is not technically correct since only $\edis$ provides a correlation length and disorder strength, while in principle, the characterizations could be distinct for the edge and bulk. A more accurate form would be $\sigma(\br,\br') = \edis(\br-\br') 1\{0\le X < \text{UV cutoff}\}+\bdis(\br-\br') \Theta(X)$ where the UV cutoff is on the length scale of the lattice constant $a$ (characterizes the width of the edge), so that the first term provides the dominant edge contribution $\delta \edge$ and the second provides the bulk contribution $\delta \bulk$.}.
\end{enumerate}
In either case, the expression for $\overline{|T(\bk',\bk)|^2}$, and hence for $\bar{J}_1$ is expressible as the sum of a term proportional to $|t_\perp|^2$
and the second term which is proportional to $\sigma$ (or its Fourier transform $\Sigma$) which reflects the presence of disorder:
\begin{equation}
    \bar{J}_1 =(\bulk +\delta\bulk)\times A + (\edge +\delta \edge)\times L+O(1)
\end{equation}
Where $\bulk,\edge$ are the values from the clean limit and $\delta \bulk,\delta \edge$ originate from the disorder correlation.

\subsubsection{Review: Bulk with Disorder}

When considering the bulk disorder, we have 
\begin{equation}
    \label{app-eq:bulk-disorder}
    \Bdis(\bq) = \frac{1}{L^2} \sum_{\br} \bdis (\br) e^{-i\bq\br}
\end{equation}
and thus
\begin{align}
    \delta \bulk \times A +O(1) &= \frac{1}{L^2} \sum_{\bk',\bk\in \dB\dZ} \Bdis(\bk-\bk') f(\bk,\bk-\bk') \\
    &=\sum_{\bq\in \dB\dZ} \Bdis(\bq) j(\bq) +O\left(1\right)\\
    &=A\times \iint_{\bq\in \dB\dZ} \Bdis(\bq)j(\bq) \frac{d^2\bq}{(2\pi)^2} +O\left(1\right)\\
    \delta \bulk &= \iint_{\bq\in \dB\dZ} \Bdis(\bq)j(\bq) \frac{d^2\bq}{(2\pi)^2}
\end{align}
Where the second equality uses Eq. \eqref{app-eq:j-def} and that $\bdis(\br) = O(1)$ (i.e., $\Bdis(\bq) = O(1)$).

If $\bdis(\br)$ has a characteristic correlation length $\xi_\text{b}$ (e.g., $\bdis(\br)$ is of Gaussian form), then its Fourier transform has correlation length $1/\xi_\text{b}$. 
If $1/\xi_\text{b}$ is smaller than the distance $\delta k_F$ between the FSs so that even after momentum boost, the two FS do not intersect for $|\bq| < 1/\xi_\text{b}$. 
Therefore, when so long as the result does not vanish by   symmetry (i.e. under the same constraints that apply in the clean limit), the disorder contribution scales like 
\begin{equation}
    \delta \bulk \sim   \frac{g_\text{b}^2|\Delta|}{E_F^2}\times\left(\frac {|\Delta|} {E_F}\right),
\end{equation}
where $g_\text{b}^2 \sim \Bdis(\br)\sim \Bdis(\bq)$ is the characteristic disorder strength (replaces the tunneling parameter $t_\perp$ in Eq. \eqref{eq:bulk-scale}).
In all other scenarios, the extra suppression factor vanishes so that
\begin{equation}
    \delta \bulk \sim   \frac{g_\text{b}^2|\Delta|}{E_F^2}\times\left(\frac {|\Delta|} {E_F}\right)^{\delta_\text{b}}
\end{equation}
Where $\delta_\text{b} = 1$ if the FS are non-intersecting ($\delta=1$, i.e., $\delta k_F \gtrsim |\Delta|/E_F$) and $\delta k_F \gtrsim 1/\xi_\text{b}$.

\subsubsection{Edge with Disorder}
\label{app:edge-disorder-correlation}

When consider an edge, it is natural to model the disorder \footnote{Although the current form of disorder $\sigma(\br,\br')$ gives both a bulk $\delta \bulk$ and edge $\delta \edge$ contribution, it is not technically correct since only $\edis$ provides a correlation length and disorder strength, while in principle, the characterizations could be distinct for the edge and bulk. A more accurate form would be $\sigma(\br,\br') = \edis(\br-\br') 1\{0\le X < \text{UV cutoff}\}+\bdis(\br-\br') \Theta(X)$ where the UV cutoff is on the length scale of the lattice constant $a$ (characterizes the width of the edge), so that the first term provides the dominant edge contribution $\delta \edge$ and the second provides the bulk contribution $\delta \bulk$.} via
\begin{equation}
    \sigma(\br,\br') = \edis(\br-\br')\Theta(X), \quad X = \frac{x+x'}{2}
\end{equation}
Note that the variance $\sigma(\br,\br')$ should be $L$-periodic with respect to $\br,\br'$. 
However, as currently defined, the average position $X\mapsto X+L/2$ under translation $x \mapsto x+L$ and $x'\mapsto x'$, and thus we need to define the variance more carefully.
The simplest way is to define $[x]$ as the $L$-modulo of the $x$ position so that $[x] \in \dZ$ and within the interval $-\ell \le [x] < \ell$ (where continuing the notation in Appendix \eqref{app:edge-comp} and taking $L=2\ell$), and replace $\Theta(X)$ with
\begin{equation}
    \Theta \left(\frac{[x]+[x']}{2} \right)\equiv 1\left\{0\le \left(\frac{[x]+[x']}{2} \in \dR \right)<\ell\right\}
\end{equation}
With this modification, one can compute the Fourier transform, i.e.,
\begin{equation}
    \Sigma(\bq) = \frac{1}{L^4} \sum_{\br',\br} \edis (\br-\br') e^{-i\bq(\br'-\br)} \Theta\left(\frac{[x']+[x]}{2} \right)
\end{equation}
Where the summation $\br,\br'$ is over $\{-\ell,...,\ell-1\}^2$, respectively.
Since the integrand is $L$-periodic with respect to $\br,\br'$, we have
\begin{align}
    \Sigma(\bq) &= \frac{1}{L^3} \sum_{\delta \br \equiv (\delta x,\delta y)\in \{-\ell,...,\ell-1\}^2 } \edis (\delta\br) e^{-i\bq\delta\br} \sum_{-\ell \le (x\in \dZ) <\ell} \Theta\left(\frac{[x+\delta x]+[x]}{2} \right) \\
    &= \frac{1}{L^3}\sum_{\delta \br } \edis (\delta\br) e^{-i\bq\delta\br} (\ell -1\{\delta x\; \text{odd} \}) \\
    &= \frac{1}{2} \Edis(\bq) - \frac{1}{2L} \underbrace{\left[ \frac{2}{L^2} \sum_{\delta \br} \edis (\delta\br)1\{\delta x\; \text{odd} \}e^{-i\bq\delta\br} \right]}_{= \Sigma_\text{e}(\bq) +O(1/L^2)}
\end{align}
where $\{\delta x \; \text{odd}\}$ is short for summing only over odd integers $\delta x$ within the interval $\delta x\in \{-\ell,...,\ell-1\}$.
Note the first term will provide the bulk disorder $\delta \bulk$ (compare with Eq. \eqref{app-eq:bulk-disorder}), while the second term will contribute to the edge disorder $\delta \edge$. 
More specifically,
\begin{align}
     \delta \bulk \times A +\delta \edge \times L+O(1) &=  \frac{1}{L^2}\sum_{\bk',\bk\in \dB\dZ} \Sigma(\bk-\bk') f(\bk,\bk-\bk') \\
    &=\sum_{\bq\in \dB\dZ} \Sigma(\bq) j(\bq) +O\left(1\right)\\
    &=A\times \iint_{\bq\in \dB\dZ} \Edis(\bq)j(\bq) \frac{d^2\bq}{(2\pi)^2} +L\times \frac{1}{2}\iint_{\bq\in \dB\dZ} \Edis(\bq)j(\bq) \frac{d^2\bq}{(2\pi)^2} +O\left(1\right)
\end{align}
Where we used the fact that the ``effective area" of a half-plane is $A=L^2/2$. Hence,
\begin{equation}
    \delta \bulk = 2\delta \edge = \iint_{\bq\in \dB\dZ} \Edis(\bq)j(\bq) \frac{d^2\bq}{(2\pi)^2}
\end{equation}
Using a similar argument for the bulk disorder, we find that
\begin{equation}
    \delta \bulk, \delta \edge \sim   \frac{g_\text{e}^2|\Delta|}{E_F^2}\times\left(\frac {|\Delta|} {E_F}\right)^{1\{\delta k_F \gtrsim 1/\xi_\text{e}, |\Delta|/E_F\}}
\end{equation}

\revise{
\section{Full Generality}
\label{app:general}
In this section, we shall consider the general case where only TRS $\sT$ is assumed, so that the result discussed in the introduction holds for multi-band systems, singlet-triplet mixing, spin-orbit coupling and so on.
}
\revise{
\subsection{Setup}
\label{app:setup}
More concretely, consider the following Hamiltonian describing the independent top/bottom SCs defined on a two-dimensional Bravais lattice,
\begin{equation}
    \sH_0^\pm = \sum_{\br,\br',s,s'} \left[(c^\pm_{\br s})^\dagger \ve_{\br s, \br' s'}^\pm c_{\br s}^\pm +\frac{1}{2} \left((c^\pm_{\br s})^\dagger \Delta^\pm_{\br s, \br' s'} (c^\pm_{\br, \sT s})^\dagger + \text{h.c.}\right)\right] 
\end{equation}
Where $\br$ denote the lattice site and $s$ denotes the possible bands (where we implicitly treat spin as a subset of the band index), and  $\sT$ denote time-reversal (anti-unitary) symmetry, so that within this representation, a single-band singlet SC is given by $\Delta_{\uparrow,\uparrow} = \Delta_{\downarrow,\downarrow}$ 
Then we can write the Hamiltonian as (ignoring unimportant constants)
\begin{equation}
    \sH_0 \equiv \sH_0^+ +\sH_0^- = \frac{1}{2} \psi^\dagger H_0 \psi
\end{equation}
where $H_0 = H_0^+ \oplus H_0^-$ and
\begin{equation}
    \psi = 
    \begin{bmatrix}
        (c^+)^\dagger \\
        c^+_\sT \\
        (c^-)^\dagger \\
        c^-_\sT
    \end{bmatrix}, \quad
    H_0^\pm =
    \begin{bmatrix}
        \ve^\pm & \Delta^\pm \\
        (\Delta^\pm)^\dagger & -\sT \ve^\pm \sT^\dagger 
    \end{bmatrix}
\end{equation}
In particular, in real $\br$ and momentum $\bk$ space, we have
\begin{equation}
    \psi = 
    \begin{bmatrix}
        (c^+_{\br s})^\dagger \\
        c^+_{\br, \sT s} \\
        (c^-_{\br s})^\dagger \\
        c^-_{\br, \sT s}
    \end{bmatrix}
    =
    \begin{bmatrix}
        (c^+_{\bk s})^\dagger \\
        c^+_{-\bk, \sT s} \\
        (c^-_{\bk s})^\dagger \\
        c^-_{-\bk, \sT s}
    \end{bmatrix}
\end{equation}
where we used the fact that $\sT$ maps $\bk \mapsto -\bk$.
Note that if $\sX,\sY,\sZ$ denotes the corresponding Pauli matrices with respect to single particle-hole Hilbert space, then $\sC \equiv \sX\sT$ denotes the particle-hole (anti-unitary) symmetry and that $H_0^\pm$ are particle-hole symmetric, i.e., $\sC H_0^\pm =-H_0^\pm \sC$.

Similarly, we can define the tunneling Hamiltonian between the two layers as
\begin{align}
    \sH_T = \sum_{\br,\br,s,s'} \left[(c^+_{\br s})^\dagger T_{\br s, \br' s'} c_{\br s}^- +\text{h.c.} \right] = \frac{1}{2}\psi^\dagger H_T \psi
\end{align}
Where 
\begin{equation}
    H_T = 
    \begin{bmatrix}
        0 & \bra +|H_T|-\ket\\
        \bra -|H_T|+\ket & 0
    \end{bmatrix}, \quad
    \bra +|H_T|-\ket =
    \begin{bmatrix}
        T & 0\\
        0 & -\sT T\sT^\dagger
    \end{bmatrix}
\end{equation}
So that $H_T$ is an inter-layer interaction and $T$ represents the tunneling operator mapping within the particle subspace (and conversely, $\sT T\sT^\dagger$ denotes the mapping within the hole subspace).
The full Hamiltonian is then given by
\begin{equation}
    \sH = \sH_0 + \sH_T = \frac{1}{2} \psi^\dagger H \psi, \quad H = H_0 +H_T
\end{equation}
If there exists a global phase difference $\alpha$ across the SC junction, the Hamiltonian is modified by a gauge symmetry, i.e.,
\begin{equation}
    \sH(\alpha) = \sH_0^+(\alpha) +\sH_0^- + \sH_T 
\end{equation}
Where
\begin{equation}
    \sH_0^+ (\alpha) =e^{i\alpha \sZ/2} \sH_0^+ e^{-i\alpha \sZ/2}
\end{equation}
So that $\Delta^+ \mapsto e^{i\alpha} \Delta^+$, while $\Delta^-$ remains the same.

The Josephson energy at zero temperature is then given by the ground state energy \begin{equation}
    \bra \sH(\alpha)\ket = \tr (H(\alpha) \dnum\{H(\alpha) <0\})
\end{equation}
Where we used the fact that $\sH(\alpha)$ denotes non-interacting quasi-particles and $\dnum\{H(\alpha) <0\}$ is the projection operator onto the the negative energy values of $H(\alpha)$.
}

\revise{
\subsection{Perturbation Theory}
If we treat the tunneling Hamiltonian as a perturbation, i.e., write $H_T = t_\perp \hat{H}_T$ where $t_\perp$ denotes the tunneling scale, then the full Hamiltonian is given by
\begin{equation}
    H_{t_\perp}(\alpha) = H_0(\alpha) + t_\perp \hat{H}_T  
\end{equation}
For sufficiently small $t_\perp$, perturbation theory (see end of subsection) tells us that up to $O(t_\perp^4)$ error,
\begin{align}
    2\bra \sH_{t_\perp}(\alpha) \ket &= \tr (H_{t_\perp}(\alpha) \dnum\{H_{t_\perp}(\alpha) <0\}) \\
    &= \tr (H_0 \dnum\{H_0 <0\} ) \\
    &\quad\quad+ t_\perp^2 \sum_{E_i <0 <E_j} \frac{\tr\left(\dnum\{H_0(\alpha) =E_i\}\hat{H}_T \dnum\{H_0(\alpha) =E_j\} \hat{H}_T \dnum\{H_0(\alpha) =E_i\} \right)}{E_i -E_j}
    \label{app-eq:pert}
\end{align}
Where the extra factor of $2$ originates from the definition of $\sH = \psi^\dagger H \psi /2$.
Note that the first term represents the ground state energy of the independent SC layers, i.e., $\bra \sH_0 \ket_0$ where $\bra \cdots \ket_0$ is the ground state with respect to $\sH_0$, and thus is independent of $\alpha$. 
The term proportional to $t_\perp^2$ thus contains the lowest order contribution to a nontrivial $\alpha$ dependence and thus will contain the first order Josephson coupling $-J_1 \cos \alpha$.
In particular, $\dnum\{H_0(\alpha) =E_i\}$ denotes the projection operator onto the eigenspace with energy $E_i$ of the free Hamiltonian $H_0 (\alpha)$. Note that the energies $E_i$ are independent of $\alpha$, but the projections operators are dependent on $\alpha$. The summation is then over all distinct energies $E_i,E_j$ such that $E_i < 0 <E_j$.
}

\revise{
By particle-hole symmetry $\sC$, we can partition the eigenspaces of $H_0^\pm$ into those with positive energy $E^\pm>0$ and corresponding projection operator $\dnum\{H_0^\pm = E^\pm\}$, and those with negative energy $-E^\pm$ and corresponding projection operator $\dnum\{H_0^\pm = -E^\pm\} =\sC \dnum\{H_0^\pm = E^\pm\} \sC^\dagger$.
Also note that since $H_0^+(\alpha)$ differs from $H_0^+$ by a gauge transform,  the corresponding eigenspaces of $H_0^+(\alpha)$ are given by the gauge-transformed projection operators $e^{i\alpha \sZ/2} \dnum\{H_0^+ = E^+\} e^{-i\alpha \sZ/2}$.
We can then rewrite Eq. \eqref{app-eq:pert} as
\begin{align}
    \eqref{app-eq:pert} &= 2\sum_{E^+,E^->0} \frac{\tr\left(\dnum\{H_0^+(\alpha) =-E^+\}H_T \dnum\{H_0^- =E^-\} H_T^\dagger \dnum\{H_0^+(\alpha) =-E^+\} \right)}{-E^+ -E^-} \\
    &= -2\sum_{E^+,E^->0} \frac{\tr\left( \dnum\{H_0^+ =E^+\} \sC^\dagger e^{-i\alpha \sZ/2} H_T \dnum\{H_0^- =E^-\} H_T^\dagger e^{i\alpha \sZ/2} \sC\dnum\{H_0^+ =E^+\}  \right)}{E^+ +E^-} \\
    &= -2\sum_{E^+,E^->0} \frac{\norm{ \dnum^+ \sC^\dagger e^{-i\alpha \sZ/2} H_T \dnum^-  }_2^2}{E^+ +E^-}
\end{align}
Where we used the fact that $H_T$ is an inter-layer interaction and the extra factor of $\times 2$ originates from the degenerate super-exchange process, i.e., 
\begin{equation}
    \dnum\{H_0^-(\alpha) =-E^-\}H_T^\dagger \dnum\{H_0^+ =E^+\} H_T \dnum\{H_0^+(\alpha) =-E^-\}
\end{equation}
We also simplify notation by writing $\dnum^\pm =\dnum\{H_0^\pm =E^\pm\}$ and $\norm{A}_2^2 = \tr (AA^\dagger)$ for the $L^2$-norm of operators.
}

\revise{
Let $\dnum_p,\dnum_h = (I\pm \sZ)/2$ denote the projection operator onto the particle, hole subspace, so that $\bra +|H_T|-\ket = \dnum_p T\dnum_p -\dnum_h \sT T\sT^\dagger \dnum_h$. Then
\begin{align}
    \eqref{app-eq:pert} &= -2 \sum_{E^+,E^->0} \frac{\norm{ e^{i\alpha/2} \dnum^+ \dnum_h \sT^\dagger T \dnum_p \dnum^-  -e^{-i\alpha/2} \dnum^+ \dnum_p T\sT^\dagger \dnum_h \dnum^-}_2^2}{E^+ +E^-} \\
    &= 2\left[ e^{i\alpha} \sum_{E^+,E^->0} \frac{\tr((\dnum_p\dnum^+ \dnum_h) \sT^\dagger T (\dnum_p \dnum^- \dnum_h) \sT T^\dagger)}{E^+ +E^-} +\text{h.c.}\right] +\cdots
    \label{app-eq:pert-final}
\end{align}
Where $\cdots$ denotes terms independent of the phase difference $\alpha$.

\begin{proof}[Proof of Eq. \eqref{app-eq:pert}]
The remainder of this subsection will be devoted to proving Eq. \eqref{app-eq:pert}.
To see its application towards symmetry considerations, skip to the next subsection.
For the sake of notation simplicity, we shall omit the $\alpha$ notation when writing $H_0(\alpha)$ within this proof.
The main goal is thus to write $\dnum\{H_{t_\perp} < 0\}$ as a series expansion using $t_\perp$.

Let $\Gamma$ be a contour in $\dC$ that encloses all negative energies of $H_0$ and no positive energies.
Then for sufficiently small $t_\perp$, no energy of $H_{t_\perp}$ crosses the contour $\Gamma$. Hence, we can define the following is well-defined
\begin{equation}
    \oint_{\Gamma} \frac{1}{z-H_{t_\perp}} \frac{dz}{2\pi i}
\end{equation}
By complex analysis, it's straightforward to check that
\begin{equation}
    \dnum\{ H_{t_\perp}<0\} = \oint_{\Gamma} \frac{1}{z-H_{t_\perp}} \frac{dz}{2\pi i}
\end{equation}
Since the right-hand-side can be written as a Taylor series expansion of $t_\perp$, it's straightforward to check that Eq. \eqref{app-eq:pert} follows.
For example,
\begin{align}
    t_\perp \frac{d}{dt_\perp}\dnum\{ H_{t_\perp}<0\} &= \oint_\Gamma \frac{1}{z-H_{t_\perp}} H_T \frac{1}{z-H_{t_\perp}} \frac{dz}{2\pi i} \\
    t_\perp \left. \frac{d}{dt_\perp}\right|_{t_\perp =0} \dnum\{ H_{t_\perp}<0\}
    &= \sum_{E_i,E_j} \dnum \{H_0=E_i\} H_T \dnum\{H_0=E_j\} \oint_\Gamma \frac{1}{(z-E_i)(z-E_j)} \frac{dz}{2\pi i} \\
    &= \sum_{E_i \ne E_j} \frac{\dnum\{H_0=E_i\} H_T \dnum\{H_0=E_j\}}{E_i-E_j} [1\{E_i <0\} -1\{E_j >0\}] \\
    &\quad\quad + \sum_{E_i \ne E_j} \frac{\dnum\{H_0=E_i\} H_T \dnum\{H_0=E_j\}}{E_i-E_j} [1\{E_i >0\} -1\{E_j <0\}] \\
    &= \sum_{E} \left[\dnum\{H_0 =E\} H_T R_0(E) + \text{h.c.}\right], \quad R_0(E) = \frac{1}{E-H_0} \dnum\{H_0 =E\}^\perp
\end{align}
Where $\dnum\{H_0 =E\}^\perp$ is the orthogonal projection and the summation in the last equality is over all eigen-energies $E$ of $H_0$. In fact, it's straightforward to check that
\begin{equation}
    \frac{t_\perp^n}{n!} \left.\left(\frac{d}{dt_\perp}\right)^n\right|_{t_\perp =0} \dnum\{H_{t_\perp} < 0\} = \sum_{E} \sum_{i=0}^n (R_0(E) H_T)^i \dnum\{H_0 =E\} (H_T R_0(E))^{n-i}
\end{equation}
In this case, we can then Taylor expand $\tr (H_{t_\perp} \dnum\{H_{t_\perp}< 0\})$ in series of $t_\perp$. For example,
\begin{align}
    t_\perp \left.\frac{d}{dt_\perp}\right|_{t_\perp =0}\tr (H_{t_\perp} \dnum\{H_{t_\perp}< 0\}) &= \tr(H_T \dnum\{H_0 <0\}) + \underbrace{\tr \left(H_0 t_\perp \left.\frac{d}{dt_\perp}\right|_{t_\perp =0}\dnum\{ H_{t_\perp}<0\}\right)}_{=0} \\
    &= \tr(H_T \dnum\{H_0 <0\})
\end{align}
Where we used the commutation relation of the trace and that $R(E) \dnum\{H_0 =E\} =0$. 
More generally, it's straightforward to check that
\begin{align}
    \frac{t_\perp^n}{n!} \left. \left(\frac{d}{dt_\perp}\right)^n \right|_{t_\perp = 0}\tr (H_{t_\perp} \dnum\{H_{t_\perp}< 0\}) &= \frac{1}{n} \tr\left(H_T  \frac{t_\perp^{n-1}}{(n-1)!} \left.\left(\frac{d}{dt_\perp}\right)^{n-1}\right|_{t_\perp =0} \dnum\{H_{t_\perp} < 0\} \right) \\
    &= \sum_{E} \tr \left(\dnum\{H_0 =E\} \underbrace{H_T R_0(E) H_T \cdots R_0 (E) H_T}_{H_T \text{ appears } n \text{ times}} \right)
\end{align}
In particular, note that $H_T$ is an inter-layer interaction and thus if $n$ is odd, then the right-hand-side $=0$ and thus the Josephson coupling only has even order contributions from $t_\perp$ (the super-exchange must alternate between the two layers, but eventually return to the same layer it started).
\end{proof}
}

\revise{
\subsection{Time-reversal Symmetry $\sT$}
We emphasize that up to this point, the derivation is general and does not use the fact that the system preserves TRS $\sT$.
TRS is only necessary so that the ground state energy/Josephson energy is invariant under $\alpha \mapsto -\alpha$.
Indeed, by TRS, it's straightforward to check that $\bra \sH(\alpha) \ket =\bra \sH(-\alpha)\ket$ and thus we can expand the term in series of cosines, i.e.,
\begin{equation}
    \bra \sH(\alpha)\ket = -J_0 -J_1 \cos \alpha -\cdots
\end{equation}
Moreover, by TRS, the coefficients of $e^{\pm i\alpha}$ in Eq. \eqref{app-eq:pert-final} are real, and thus the perturbation formula simplifies to
\begin{align}
    \eqref{app-eq:pert-final} = \cos \alpha\times 4\sum_{E^+,E^->0}\frac{\tr((\dnum_p\dnum^+ \dnum_h) T (\dnum_p \dnum^- \dnum_h)  T^\dagger)}{E^+ +E^-} +\cdots
\end{align}
In particular, up to higher orders $O(t_\perp^4)$,
\begin{equation}
    \label{app-eq:J1-general}
    J_1 = -2\sum_{E^+,E^->0}\frac{\tr((\dnum_p\dnum^+ \dnum_h) T (\dnum_p \dnum^- \dnum_h)  T^\dagger)}{E^+ +E^-}
\end{equation}
}

\revise{

\subsection{Momentum Space Representation}
Since the free Hamiltonian $H_0$ is translation invariant, it's often convenient to write Eq. \eqref{app-eq:J1-general} in momentum space $\bk$. 
More concretely, $H_0^\pm$ are local in momentum space $\bk$ and thus we can write 
\begin{equation}
    H_0^\pm =\bigoplus_{\bk \in \dB\dZ} H_0^\pm (\bk)
\end{equation}
By particle-hole symmetry $\sC$, we can partition the eigenstates of $H_0^\pm$ into those with positive energy, denoted by $|\bk\ell\ket^\pm$ with energy $E_{\bk\ell}^\pm >0$, and those with negative energy, denoted by $\sC\left|-\bk, \ell\right\ket^\pm$ (note $\sC$ maps $-\bk \mapsto \bk$) with energy $-E_{-\bk\ell}^\pm$.
In particular, the  quasi-particle creation operator corresponding to $|\bk\ell\ket^\pm$ is given by
\begin{equation}
    (\psi_{\bk \ell}^\pm)^\dagger = \sum_{s} \left[u_{\bk;\ell,s}^\pm (c^\pm_{\bk,s})^\dagger +v_{\bk;\ell,s}^\pm c^\pm_{-\bk,\sT s}\right], \quad u_{\bk;\ell,s}^\pm,v_{\bk;\ell,s}^\pm \in \dC
\end{equation}
Note that if there are $S$ bands in total, then $\ell=1,2,...,S$. 
The Josephson coupling in Eq. \eqref{app-eq:J1-general} can then be rewritten as
\begin{equation}
    \label{app-eq:J1-general-k}
    J_1 = -2\sum_{\bk\ell,\bk'\ell'}\frac{\tr((\dnum_p\dnum_{\bk\ell}^+ \dnum_h) T (\dnum_p \dnum^-_{\bk'\ell'} \dnum_h)  T^\dagger)}{E^+_{\bk\ell} +E^-_{\bk'\ell'}}
\end{equation}
Where $\dnum_{\bk\ell}^\pm$ denotes the projection operator onto eigenstate $|\bk\ell\ket^\pm$.

If the two layers were additionally single-band singlet SCs with a $SU(2)$-invariant tunneling matrix $T$ and normal state dispersion $\ve^\pm$, then Eq. \eqref{app-eq:J1-general-k} reduces to Eq. \eqref{app-eq:J1-singlet} (up to unimportant constants).
More concretely, we can choose
\begin{equation}
    (\psi_{\bk s}^\pm)^\dagger = \sqrt{1+\frac{\ve_{\bk}^\pm}{2E_{\bk}^\pm}}(c^\pm_{\bk,s})^\dagger +\frac{\Delta_{\bk}^\pm}{|\Delta_{\bk}^\pm|}\sqrt{1-\frac{\ve_{\bk}^\pm}{2E_{\bk}^\pm}} c^\pm_{-\bk,\sT s}
\end{equation}
}
\revise{
\subsection{General Rotation}

In Appendix \eqref{app:setup}, we have assumed that the system possesses an underlying Bravais lattice.
This can only be true provided that the two layers are ideally aligned. 
Therefore, to include general twist angles, one would fix the tunneling matrix $T$, and replace $H_0^\pm \mapsto \sR^\pm H_0^\pm (\sR^\pm)^\dagger$ (and corresponding $\dnum^\pm \mapsto \sR^\pm \dnum^\pm (\sR^\pm)^\dagger$) where $\sR^\pm$ corresponds to a rotation of the top/bottom layers by $\phi^\pm$.
Hence, the Josephson coupling at arbitrary twist angles $\phi^\pm$ is given by
\begin{equation}
    \label{app-eq:J1-general-rotate}
    J_1(\phi^\pm) = -2\sum_{E^+,E^-}\frac{\tr((\dnum_p\dnum^+ \dnum_h) (\sR^+)^\dagger T \sR^- (\dnum_p \dnum^- \dnum_h)  (\sR^-)^\dagger T^\dagger \sR^+)}{E^+ +E^-}
\end{equation}
For example, in the half-plane setup described in Fig. \ref{fig:setup}, the previous description corresponds to a coordinate system with the edge along the principle axis with rotation $\phi^\pm =\phi, \theta+\phi$.
Alternatively, if the two layers are full-planes so that the tunneling matrix $T$ is (approximately) invariant under $SO(2)$ rotation, then the previous equation implies that $J_1(\phi^\pm)$ only depends on the relative twist $\theta = \phi^- -\phi^+$, as expected for a bulk contribution.
}

\revise{
\subsection{Symmetry Considerations}
Suppose that $\sS$ is a point group symmetry which maps $\ve^\pm ,\Delta^\pm \mapsto \ve^\pm, s^\pm \Delta^\pm$ with $s^\pm \in \{ \pm 1\}$, and commutes with particle, hole projections $\dnum_p,\dnum_h$ and TRS $\sT$. 
We claim that (see end of subsection for proof) for any energy $E^\pm$ of $H_0^\pm$,
\begin{equation}
    \label{app-eq:symmetry}
    \sS \left(\dnum_p \dnum^\pm  \dnum_h \right)\sS^\dagger = s^\pm \left(\dnum_p \dnum^\pm \dnum_h \right)
\end{equation}
Hence, by Eq. \eqref{app-eq:J1-general-rotate}, if $\sS$ denotes rotation, then $J_1 (\phi^\pm) \mapsto J_1(\sS\phi^\pm)$ corresponds to replacing $\sR^\pm \mapsto \sR^\pm \sS =\sS \sR^\pm$ so that
\begin{align}
    \sR^\pm \dnum^\pm (\sR^\pm)^\dagger&\mapsto \sR^\pm \sS \dnum^\pm \sS^\dagger (\sR^\pm)^\dagger \\
    \dnum_p \sR^\pm \dnum^\pm (\sR^\pm)^\dagger \dnum_h &\mapsto s^\pm \dnum_p \sR^\pm \dnum^\pm (\sR^\pm)^\dagger \dnum_h \\
    J_1(\phi^\pm) &\mapsto s^+ s^- J_1(\phi^\pm) \\
    J_1(\sS\phi^\pm) &= s J_1(\phi^\pm), \quad s=s^+s^-
\end{align}
Moreover, if $\sS$ is also a symmetry of the tunneling matrix $T\mapsto T$. 
Then we see that $J_1(\sS \phi^\pm) = J_1(\phi^\pm)$ and thus if $s=-1$, then $J_1(\phi^\pm) =0$ for all angles $\phi^\pm$.
This corresponds to the bulk scenario where $T$ is invariant under $C_{2n}$ rotation. 
However, if $T$ corresponds to an edge/half-plane, then the edge/half-plane breaks rotation symmetry and thus $J_1(\phi^\pm)$ is generally non-vanishing.

Alternatively, if $\sS$ denotes mirror symmetry, then $J_1 (\phi^\pm) \mapsto J_1(\sS\phi^\pm)$ corresponds to replacing $\sR^\pm \mapsto \sR_{\sS \phi^\pm} = \sS \sR^\pm \sS$ so that
\begin{align}
    \sR^\pm \dnum^\pm (\sR^\pm)^\dagger&\mapsto \sS \sR^\pm \sS \dnum^\pm \sS^\dagger (\sR^\pm)^\dagger \sS^\dagger \\
    \dnum_p  \sR^\pm \dnum^\pm (\sR^\pm)^\dagger \dnum_h &\mapsto s^\pm \sS\dnum_p \sR^\pm \dnum^\pm (\sR^\pm)^\dagger \dnum_h \sS^\dagger 
\end{align}
Using Eq. \eqref{app-eq:J1-general-rotate}, we see that if $\sS$ is also a symmetry of the tunneling matrix, i.e., $T\mapsto T$, then
\begin{equation}
    J_1(\sS \phi^\pm) = s J_1(\phi^\pm)
\end{equation}
Since we have not assumed the detailed structure of the tunneling matrix $T$, this result holds for both bulk and edge contributions (e.g., if $T$ corresponds to an edge/half-plane, then mirror symmetry along the edge would keep the tunneling matrix invariant).
}

\revise{
\begin{proof}[Proof of Eq. \eqref{app-eq:symmetry}]
The remainder of this subsection is devoted to proving Eq. \eqref{app-eq:symmetry}.
First note that
\begin{align}
    \sS \dnum_p H^\pm_0 \dnum_h \sS^\dagger &=  \sS \Delta^\pm \sS^\dagger  =s^\pm \dnum_p H^\pm_0 \dnum_h\\
    \sS \dnum_p (H^\pm_0)^2  \dnum_h\sS^\dagger &=  \sS (\ve^\pm \Delta^\pm -\Delta^\pm \sT \ve^\pm  \sT^\dagger )\sS^\dagger  =s^\pm \dnum_p (H^\pm_0)^2 \dnum_h
\end{align}
Continuing this process and we see that for any polynomial $f(H^\pm_0)$, we have
\begin{equation}
    \sS \dnum_p f(H^\pm_0) \dnum_h \sS^\dagger = s^\pm  \dnum_p f(H^\pm_0) \dnum_h
\end{equation}
Using continuity arguments, one can extend the equality to all ``well-behaved" functions $f$. 
One particular important function is the resolvent, i.e.,
\begin{equation}
    \frac{1}{z-H^\pm_0}, \quad z\in \dC
\end{equation}
If $\Gamma^\pm$ denotes a contour in $\dC$ that encloses only $E^\pm$ and no other energy of $H_0^\pm$, then by complex analysis, it's straightforward to check that
\begin{equation}
    \dnum^\pm = \oint_{\Gamma^\pm} \frac{1}{z-H^\pm_0}\frac{dz}{2\pi i}
\end{equation}
Hence, the claim follows.
\end{proof}
}

\revise{
\subsection{Disorder}
\label{app:general-disorder}
Similar to the clean system, the symmetry considerations of disorder can also be proven generally.
More concretely, the tunneling matrix acts as a quenched disorder operator so that the formula of $J_1$ in Eq. \eqref{app-eq:J1-general} is replaced by
\begin{equation}
    \label{app-eq:J1-general-disorder}
    J_1 = -2\sum_{E^+,E^->0}\frac{\overline{\tr((\dnum_p\dnum^+ \dnum_h) T (\dnum_p \dnum^- \dnum_h)  T^\dagger)}}{E^+ +E^-}
\end{equation}
Where $\overline{\cdots}$ denotes disorder-averaging. If we write $T= \bar{T} +\delta T$ so that $\bar{T}$ is the disorder-average, then it's straightforward to check that the clean and disorder contributions to the Josephson coupling decouple in the sense that
\begin{align}
    J_1 = &\underbrace{-2\sum_{E^+,E^->0}\frac{\tr((\dnum_p\dnum^+ \dnum_h) \bar{T} (\dnum_p \dnum^- \dnum_h)  \bar{T}^\dagger)}{E^+ +E^-}}_{\text{clean}} \\
    &\underbrace{-2\sum_{E^+,E^->0}\frac{\tr\left[ ((\dnum_p\dnum^+ \dnum_h) \otimes (\dnum_p \dnum^-\dnum_h)) (\overline{\delta T \otimes(\delta T)^\dagger}) \sE\right]}{E^+ +E^-}}_{\text{disorder}}
\end{align}
Where we have used the fact that $\tr(ABCD)=\tr((A\otimes C)(B\otimes D) \sE)$ where $\sE|1\ket\otimes|2\ket=|2\ket\otimes |1\ket$ is the exchange operator.
Note that the first term is the clean contribution since it only depends on the disorder-average tunneling $\bar{T}$, while the second term is the disorder contribution since it only depends on the disorder covariance $\overline{\delta T \otimes(\delta T)^\dagger}$.
The same symmetry arguments discussed in the previous subsection can then be applied to each contribution.
}